\documentclass[10pt]{article}
\usepackage{amssymb}
\usepackage{epsfig}
\usepackage{graphics}
\newtheorem{theorem}{Theorem}

\newtheorem{lemma}{Lemma}

\title{Approximation Multivariate Distribution with pair copula Using the Orthonormal Polynomial and Legendre Multiwavelets basis functions}
\author{{\textbf{ A. Daneshkhah, G. A. Parham, O. Chatrabgoun, M. Jokar} }\\
Department of Statistics, Faculty of Mathematical Sciences and Computer\\
Shahid Chamran University, Ahvaz 6135714463, Iran\\
a.daneshkhah@scu.ac.ir, parham$_{-}$g@scu.ac.ir,
o-chatrabgoun@phdstu.scu.ac.ir}

\begin{document}
\maketitle

\begin{abstract}
In this paper, we concentrate on new methodologies for copulas introduced and developed by Joe, Cooke, Bedford, Kurowica,
Daneshkhah and others on the new class of graphical models called vines as a way of constructing higher dimensional distributions. We
develop the approximation method presented by Bedford et al (2012) at which they show that any $n$-dimensional copula density can be
approximated arbitrarily well pointwise using a finite parameter set of 2-dimensional copulas in a vine or pair-copula construction. Our
constructive approach involves the use of minimum information copulas that can be specified to any required degree of precision
based on the available data or experts' judgements. By using this method, we are able to use a fixed finite dimensional family of copulas to be employed in  a vine construction, with the promise of a uniform level of approximation.
\par
The basic idea behind this method is to use a two-dimensional ordinary polynomial series to approximate any log-density of a
bivariate copula function by truncating the series at an appropriate point. We present an alternative approximation of the multivariate distribution of interest by considering orthonormal polynomial and Legendre multiwavelets as the basis functions. We show the derived approximations are more precise and computationally faster with better properties than the one proposed by Bedford et al. (2012). We then apply our method to modelling a dataset of Norwegian financial data that was previously analysed in the series of papers, and finally compare our results by them.

\end{abstract}
\section{Introduction}
Bedford and Cooke (2001, 2002) introduce a probabilistic
construction of multivariate distributions based on the simple
graphical model called \textit{vine}. This model represents an
entirely new approach of building complicated multivariate and
highly dependent models which can be seen as the classical
hierarchical modelling. The principle behind the vine construction
is to model dependency using simple local building blocs based on
conditional independence (e.g. cliques in random fields). Aas et al
(2009) called these building blocs, \textit{pair-copulae}. They use
the pair-copula decomposition of a general multivariate distribution
and propose a method to perform inference.
\par
They investigate modelling complicated high-dimensional data by fitting different parametric bivariate copulas
to construct the corresponding pair-copula model. However, there is a huge
number of parametric bivariate copulas, but it is well known that building higher-dimensional copulae is generally a difficult
problem, and choosing a parametric family for the given higher-dimensional copula is rather more difficult and limited (see
Embrechts et al., 2003). As a result, the problem of choosing a parametric copula
for a higher-dimensional copula is reduced to fitting a parametric
bivariate copulas to data. Bedford et al. (2012) stated that the use of a copula to model dependency is simply a translation of one difficult problem into another: instead of the difficulty of specifying the full joint distribution we have the difficulty of specifying the copula.
The main advantage is the technical one that copulas are normalized to have support on the
unit square and uniform marginal distributions. Therefore, the
potential flexibility of the copula, by restricting them to a
particular parametric class (e.g., Gaussian, multivariate
$t$-student, etc) is not realized in practice.
\par
To overcome this difficulty, Bedford et al (2012) proposed an alternative approach at which a vine structure can be used to approximate
any given multivariate copula to any required degree of approximation. This method can be easily implemented in practice. It is only required to assume that the multivariate copula density of interest must be continuous and non-zero.
\par
This method is constructive and involves the use of minimum information copulas that can be determined to any required degree of precision based on the available data or expert judgements. It can be shown that good approximation `locally' guarantees good approximation
globally. It can be shown hat a vine structure imposes no restrictions on the underlying joint probability distribution it
represents (Bedford et al., 2012). Furthermore, Kurowicka and Joe (2011) reported that this is essential to address this question that which vine structure is most appropriate where some structures allow the use of less complex conditional copulas than others. Conversely, if we only allow certain families of copulas then one vine structure might fit better than another. This question is still open and under study, and is beyond the scope of this paper.
\par
Thus, it is trivial to show that if there is any difficulty to fit a multivariate distribution by a pair-copulae model, then the problem
is not related to the vine structure but the copulae/conditional copulae. As a result, the question ``does a vine structure fit''
only makes sense in the context of a given family of copulae. Therefore, we need to have a class of copulae with which we can
approximate any given copula to an arbitrary degree.
\par
A natural way to build a minimum information copula or specifying
dependency constraints is through the use of moments (Bedford,
2006). These can be specified either on the copula or on the
underlying bivariate density. We follow Bedford et al. (2012) to
consider the moment constraints in which real-valued functions
$\phi_{1},\ldots,\phi_{k}$ are required to take expected values
$e_{1},\ldots,e_{k}$, respectively. We then fit a minimum
information copula that satisfies a set of constraints as above and
which has minimum information (with respect to the uniform copula
$c(u, v) = uv$) amongst the class of all copulas satisfying those
constraints. It is trivial to show that this copula is the ``most
independent'' bivariate density that satisfies these constraints. In
addition, a specification of minimum information bivariate copulas
naturally leads us to the minimum information vine distributions.
Particularly, it can be shown that if a minimal information copula
satisfied each of the (local) constraints (on moments, rank
correlation, etc.), then the resulting joint distribution would also
be minimally informative given those constraints (see Kurowicka and
Cooke, 2006).
\par
In order to calculate the minimum information copula associated with
the constraints mentioned above, an iterative numerical method
called $D_{1}AD_{2}$ algorithm is used by Bedford and Meeuwissen
(1997). The number and type of the real-valued functions
$\phi_{1},\ldots,\phi_{k}$ can control the accuracy of the
approximation approach and the cost of computation. Bedford et al
(2012) develop this method by using the ordinary polynomial bases to
approximate a multivariate distribution of interest.
\par
The main objective of this paper is to improve the density approximation proposed by Bedford et al (2012) by considering
several other bases including orthonormal polynomial series and Legendre multiwavelets, and examine their properties and possible
applications. By using orthonormal polynomial basis functions the accuracy of approximation will be increased and the computation cost
will be considerably decreased. We will show that orthonormal polynomial bases are more convenient than the other natural bases (e.g. polynomial series) for the purpose of calculation.
\par
In addition to the orthonormal polynomial bases which exhibits very nice properties and efficient to implement in practice, we can improve the approximation of a multivariate density even further using the wavelets which have been recently used for density estimation. The wavelets have become popular due to their ability to approximate a large class of functions, including those with localized, abrupt variations. However, a well-known attribute
of wavelet bases is that they can not be simultaneously symmetric, orthogonal, and compactly supported. Multiwavelets--a more general, vector--valued, construction of wavelets—-overcome this disadvantage, making them natural choices for estimating density functions, many of which exhibit local symmetries around features such as a mode. In particular, using Legender multiwavelets as basis functions will improve
accuracy of approximation incredibly and the computation cost will be considerably decreased even in comparison of the orthonormal polynomial bases. We show the efficiency of our method using the mentioned bases as above by comparing them with the model developed by Bedford et al. (2012) and the one proposed by Aas et al (2009) for modeling the Norwegian financial data which has been also studied by these authors.
\par
The paper is organised as follows. In Section 2, we introduce the
pair-copula decomposition associated with a multivariate
distribution of interest. As an example for better understanding, we
also present a vine structure regarding the Norwegian financial data
in this section. We briefly study the minimum information copula and
the approximation approach presented by Bedford et al (2012) in
Section 3. In section 4, we develop the minimum information copula
based approximation method to estimate corresponding multivariate
distribution. We develop this method using orthonormal polynomial
series (obtained based on Graham-Schmidt method) and Legender
multiwavelets as the basis functions in Section 5. In section 5, we also
illustrate how to construct Legender multiwavelets basis. In Section 6, we apply our method based on these new bases to modelling Norwegian Financial returns data. We also exhibit the potential flexibility of our approach by comparing it
with the other methods. The future directions of this work and
some other conclusions will be given in Section 7.

\section{Vine Constructions of multiple dependence}
Kurowicka and Cooke (2006) highlighted the point that however, the
copula families, such as the exchangeable multivariate Archimedean
copula or the nested Archimedean constructions, constitute a huge
improvement, but they are still not rich enough to model all
possible mutual dependencies amongst the $n$ variables. This is also
illustrated by Aas et al (2009) and Bedford et al (2012). Therefore,
a more flexible structure called \textit{pair-copula construction}
or \textit{vine} proposed by them which allows for the free
specification of $n(n - 1)/2$ copulae and is hierarchical in nature.
This modelling structure is based on a decomposition of a
multivariate density into a cascade of bivariate copulae.
\par
In other words, a vine associated with $n$ variables is a nested set
of trees, where the edges of the tree $j$ are the nodes of the tree
$j + 1;~j=1,\ldots, n-2$, and each tree has the maximum number of
edges. A \textit{regular vine} on $n$ variables is a vine in which
two edges in tree $j$ are joined by an edge in tree $j+1$ only if
these edges share a common node, $j=1,\ldots, n-2$. There are
$n(n-1)/2$ edges in a regular vine on $n$ variables. The formal
definition of vine and regular vine can be found in Kurowicka and
Cooke (2006). The following theorem expresses a regular vine distribution in terms of its density.
\par
\begin{theorem} Let $\mathcal{V} = (T_{1}, \ldots, T_{n-1})$ be a regular vine on $n$ elements,
 where $T_{1}$ is a connected tree with nodes $N_{1} = \{1, \ldots , n\}$ and edges $E_{1}$;
 for $i = 2, \ldots, n-1$, $T_{i}$ is a connected tree with nodes $N_{i} = E_{i-1}$. For each
edge $e(j, k)\in T_{i}, i = 1, \ldots, n-1$ with conditioned set $\{j, k\}$ and conditioning
set $D_{e}$, let the conditional copula and copula density be $C_{jk\mid De}$ and $c_{jk\mid De}$
 respectively. Let the marginal distributions $F_{i}$ with densities $f_{i}, i = 1, \ldots, n$ be given.
  Then, the vine-dependent distribution is uniquely determined and has a density given by
\begin{equation}
f(x_{1},\ldots,x_{n})=\prod_{i=1}^{n}f(x_{i})\prod_{j=1}^{n-1}\prod_{e(j,k)\in E_{i}}c_{jk\mid D_{e}}(F_{j\mid D_{e}}, F_{k\mid D_{e}})\label{vine-dens}
\end{equation}
\end{theorem}
\textit{Proof}. See Bedford and Cooke (2001).\\
\par
\begin{figure}
\begin{center}
\vspace{-6cm}
\scalebox{0.65}{\includegraphics{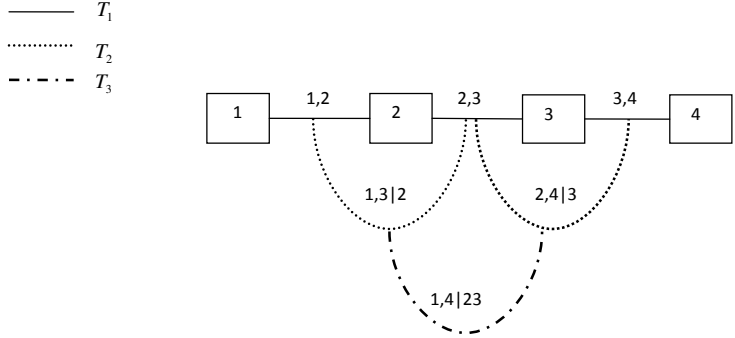}}
\end{center}
\vspace{-7.2cm}
\caption{A regular vine with 4 elements\label{vine-4vars}}
\end{figure}

The density decomposition associated with $4$ random variables $\mathbf{X}=(X_{1},\ldots,X_{4})$ with a joint
density function $f(x_{1},\ldots,x_{4})$ satisfying a copula-vine structure (this structure is called \textit{D}-vine,
see Kurowicka and Cooke, 2006, pp. 93) shown in Figure \ref{vine-4vars} with the marginal densities $f_{1},\ldots,f_{4}$ is illustrated as follows
\[
f(x_{1},\ldots,x_{4})=\prod_{i=1}^{4}f(x_{i})\times c_{12}\{F(x_{1}), F(x_{2})\} c_{23}\{F(x_{2}), F(x_{3})\}c_{34}\{F(x_{3}), F(x_{4})\}\times
\]
\begin{equation}
c_{13\mid 2}\{F(x_{1}\mid x_{2}), F(x_{3}\mid x_{2})\}c_{24\mid 3}\{F(x_{2}\mid x_{3}), F(x_{4}\mid x_{3})\}\times c_{14\mid 23}\{F(x_{1}\mid x_{2}, x_{3}), F(x_{4}\mid x_{2}, x_{3})\}\label{Vine-exam}
\end{equation}

It is trivial to show that if $f(x_{1},\ldots,x_{n})$ is absolutely continuous to product $f_{1},\ldots,f_{n}$, it
then can be represented by any vine-dependent distribution. The existence of regular vine distributions in
details is discussed in Bedford and Cooke (2002). We illustrate briefly how such a distribution is determined
using the regular vine in Figure \ref{vine-4vars} as an example. We make use of the expression
\[
f(x_{1},\ldots,x_{4})=f(x_{1})f(x_{2},\mid x_{1})f(x_{3} \mid x_{1}, x_{2})f(x_{4} \mid x_{1},\ldots,x_{3})
\]
The marginal distribution of $X_{1}$ is known, so we have $f_{1}$. The marginals of $X_{1}$ and $X_{2}$ are known, and the copula of $X_{1}$, $X_{2}$ is also known, so we can get $f(x_{1}, x_{2})$, and hence $f(x_{2}\mid x_{1})$. In order to get $f(x_{3}\mid x_{1}, x_{2})$ we can determine $f(x_{3}\mid x_{2})$ in the similar way as $f(x_{2}\mid x_{1})$. Next we calculate $f(x_{1}\mid x_{2})$ from $f(x_{1}, x_{2})$. With $f(x_{1}\mid x_{2})$, $f(x_{3}\mid x_{2})$, and the conditional copula of $X_{1}, X_{3}$ given $X_{2}$ we can determine the conditional joint distribution $f(x_{1}, x_{3}\mid x_{2})$, and hence the conditional marginal $f(x_{3}\mid x_{1}, x_{2})$. Progressing in this way we obtain $f(x_{4} \mid x_{1},\ldots,x_{3})$. As a result, we can state the following theorem.
\par
\begin{theorem}
Given a distribution with density function $f(x_{1},\ldots,x_{n})$ and a vine $\mathcal{V}$ on $n$ elements,
 there are copulae $c_{jk\mid D_{e}}$ such that (\ref{vine-dens}) is satisfied, that means
\[
f(x_{1},\ldots,x_{n})=\prod_{i=1}^{n}f(x_{i})\prod_{j=1}^{n-1}\prod_{e(j,k)\in E_{i}}c_{jk\mid D_{e}}(F_{j\mid D_{e}}, F_{k\mid D_{e}})
\]
\end{theorem}
{\bf Proof:} It is trivial, one should follow the explanation given above to build a 4-dimensional multivariate distribution to prove this theorem. See also Bedford et al. (2012) and references therein.\\
\par
The above theorem gives us a constructive approach to build a multivariate distribution
given a vine structure: If we make choices of marginal densities and copulae then the above
formula will give us a multivariate density. Hence vines can be used to model general multivariate densities.
However, in practice we have to use copulae from a convenient class,
and this class should ideally be one that allows us to approximate any given copula to an
arbitrary degree. In the following sections, we address this issue in more detail. By having
this class of copulae, we then can approximate any multivariate distribution using any vine
structure.
\par
Unlike the situation with Bayesian networks, where not all structures can be used to
model a given distribution, the theorem shows that - in principle - any vine structure may be
used to model a given distribution. However, in practice it seems that some vine structures
do work better than others, and so this must be a result of restricting to a particular family
of copulas. That is, given a family of copulae, some vine structures may give a better degree
of approximation than others. In fact, we could say that the question ``does a vine structure
fit" only makes sense in the context of a given family of copulae.

\section{Building bivariate minimum information copulae}

This section sets out to show that we can use the minimum
information techniques originated from Bedford and Meeuwissen (1997) in
conjunction with the observed data or expert elicitation of
observables, to define a copula that can be used to build the joint
distribution of two random variables. The method that will be
described below is based on using the
$D_{1}AD_{2}$ algorithm to determine the copula in terms of
potentially asymmetric information about two variables of interests.

\subsection{The $D_{1}AD_{2}$ algorithm and minimum information copula}

Bedford and Meeuwissen (1997) applied a so-called $DAD$ algorithm to
produce discretized minimally informative copula between two
variables with given rank correlation. This approach relies on the
fact that the correlation is determined by the mean of the symmetric
function $UV$. The same approach can be used whenever we wish to
specify the expectation of any symmetric function of $U$ and $V$
(see Bedford, 2006; Lewandowski, 2008).
\par
This method can be developed further using the idea stated in Borwein et al. (1994) which enables us to have
asymmetric specifications. In the revised method, we first determine a
positive square matrix $A$, also called a \textit{kernel}, and two
diagonal matrices $D_{1}$ and $D_{2}$ should be then found in such a
way that the following product, $D_{1}AD_{2}$ is doubly stochastic.
The theory can be easily generalised for continuous functions (see
Bedford et al, 2012).
\par
Now, suppose there are two random variables $X$ and $Y$, with cumulative
distribution functions $F_{X}$ and $F_{Y}$, respectively. These are
the variables of interest that we would like to correlate by
introducing constraints based on some knowledge about functions of
these variables. Suppose there are $k$ of these functions, namely
$h'_{1}(X, Y), h'_{2}(X, Y ), \ldots, h'_{k}(X, Y)$, and that we
wish either to calculate their mean values in terms of the observed
data, or the expert wishes to specify mean values
$\alpha_{1},\ldots, \alpha_{k}$ for all these functions, respectively. We can simply specify corresponding functions of the copula
variables $U$ and $V$, defined by $h_{i}(U, V) =h'_{i}(F_{1}^{-1}
(U); F_{2}^{-1} (V)),~~i=1,2,\ldots,k$, where $h_{i}: [0,1]^{2}\to \mathbb{R}$, at which we can specify the mean values $\alpha_{1},\ldots,
\alpha_{k}$ that these functions should simultaneously take. Further
suppose that $h_{i}, h_{j}$ are linearly independent for $i\neq j$.
We seek a copula that has these mean values, a problem which is
usually either infeasible or under determined. Hence, assuming
feasibility for the moment, we also ask that the copula be minimally
informative (with respect to the uniform distribution), which
guarantees a unique and reasonable solution. We form the kernel
\begin{equation}
A(u, v) = \exp(\lambda_{1}h_{1}(u, v) + \ldots + \lambda_{k}h_{k}(u, v))\label{kernel}
\end{equation}
where $u$ denote the realization of $U$ and $v$ the realization of $V$.
\par
For practical implementations, we use the same method as proposed by
Bedford et al (2012) to discretize the set of $(u, v)$ values such
that the whole domain of the copula is covered. Thus, the
aforementioned kernel $A$ becomes a 2-dimensional matrix, and two
matrices $D_{1}$ and $D_{2}$ should be then determined. As a result,
the following product denoted by $P$ over $[0, 1]^{2}$ becomes a
doubly stochastic matrix which represents a discretized copula
density.
\begin{equation}
P = D_{1}AD_{2}\label{dad}
\end{equation}
The $D_{1}AD_{2}$ algorithm can be used to generate a unique joint density with uniform marginals for each vector $(\lambda_{1},\ldots,
 \lambda_{k})$. The set of all possible expectation vectors $(\alpha_{1},\ldots, \alpha_{k})$ that could be taken by $(h_{1}, h_{2},
\ldots, h_{k})$ under some probability distribution is convex, and that for every $(\alpha_{1},\ldots, \alpha_{k})$ in the interior of that
convex set there is a density with parameters $(\lambda_{1}, \ldots ,\lambda_{k})$ for which $(h_{1}, h_{2}, \ldots, h_{k})$ take these values
 (see Borwein et al., 1994; and Bedford et al., 2012).
\par
We now explain the iterative algorithm required to approximate the
mentioned copula density by this algorithm. Suppose that both
$(u,v)$ are discretized into $n$ points, respectively as $u_{i}$,
and $v_{j},~~i,j = 1, \ldots, n$. Then, we write $A = (a_{ij}), D_{1} = diag (d_{1}^{(1)},\ldots, d_{n}^{(1)}), D_{2} = diag (d_{1}^{(2)},\ldots, d_{n}^{(2)})$,
where $a_{ij} = A(u_{i}, v_{j})$, $d_{i}^{(1)}=D_{1}(u_{i})$, $d_{j}^{(2)}=D_{2}(v_{j})$. We define the doubly stochastic matrix, $D_{1}AD_{2}$ with the uniform marginals as follows
\[
\forall i=1,\ldots n~~\sum_{j} d_{i}^{(1)}d_{j}^{(2)}a_{ij}=1/n,~~~and
\]
\[
\forall j=1,\ldots n~~\sum_{1} d_{i}^{(1)}d_{j}^{(2)}a_{ij}=1/n,
\]
The idea behind of $D_{1}AD_{2}$ algorithm is very simple which starts with arbitrary positive initial matrices for
$D_{1}$ and $D_{2}$, and the new vectors will then be successively defined by iterating the following maps
\[
d_{i}^{(1)}\mapsto\frac{1}{n\sum_{j} d_{j}^{(2)}a_{ij}}~ (i=1,\ldots,n),~~d_{j}^{(2)}\mapsto\frac{1}{n\sum_{i} d_{i}^{(1)}a_{ij}},~~(j=1,\ldots,n)
\]
It can be shown that this iteration scheme converges geometrically to the requested vectors (see Borwein et al., 1994).
\par
Note that to compare different discretizations (for different $n$) we should multiply each cell weight
$d_{i}(1)d_{j}(2)a_{ij}$ by $n^{2}$ as this quantity approximates the continuous copula density with respect to the uniform distributions.
\par
The mapping from the set of vectors of $\lambda$'s onto the set of vectors of resulting
expectations of functions $(h_{1},\ldots,h_{k})$ has to be found numerically. Bedford and Daneshkhah (2010) and Bedford et al (2012) proposed the
optimization techniques to determine the $\lambda_{i}$'s and corresponding copula. The expectations $\alpha_{i}$
of $k$ functions of variables $X$ and $Y$ are given by
\[
E[h'_{i}(X, Y)] = E[h_{i}(U, V )] = \alpha_{i},~~ i = 1,\ldots, k.
\]
We now wish to determine the appropriate set of $\lambda$'s for given  expectations $\alpha_{i}$,
where the expectations have been calculated using the discrete copula density $D_{1}AD_{2}$ given in (\ref{dad}).
Hence, to determine $\lambda_{i}$'s satisfying the constraints, the following set of equations has to be solved
\begin{equation}
L_{l}(\lambda_{1},\ldots,\lambda_{k}) =\frac{1}{n^{2}}\sum_{i=1}^{n}\sum_{j=1}^{n} P(u_{i},v_{j})h_{l}(u_{i}, v_{j})-\alpha_{l},~~~l= 1,2,\ldots,k.
\end{equation}
The left hand sides of the above equations are just functions of $\lambda$'s and with optimization
algorithms their roots can be found. One of the possible solvers for this task would be FSOLVE - MATLAB's
optimization routine. An alternative method is to use another MATLAB's optimization procedure called FMINSEARCH,
which implements the Nelder-Mead simplex method (see Lagarias et al., 1998). The minimized function is then
\[
L_{sum}(\lambda_{1},\ldots,\lambda_{k}) = \sum_{l=1}^{k} L_{l}^{2}(\lambda_{1},\ldots,\lambda_{k}).
\]
We refer the interested reader to Lewandowski (2008) and Bedford et al (2012) to show how an expert could specify a copula though defining expected values.

\section{Approximating Multivariate Density by Vine}
In this section, we use techniques from approximation theory to show
that any $n$-dimensional multivariate density which is $C^2$ (that
is, twice differentiable, with continuous second derivatives) can be
approximated arbitrarily well pointwise using a finite parameter set
of 2-dimensional copulas in a vine construction. The basic idea is
that we can use a series expansion, like a two-dimensional
Polynomial series, orthonormal Polynomial series or Legender
multiwavelets, to approximate any log-density function by truncating
the series at an appropriate point. What is non-trivial, however,
about this method, is that the same truncation can be used
everywhere in a vine construction and gives overall uniform
pointwise approximation. Hence our method allows the use of a fixed
finite dimensional family of copulas to be used in a vine
construction, with the promise of a uniform level of approximation. Since the approximations we make of copula densities might not be quite copula densities themselves, we need to transform them to make them copulas.
\par
To demonstrate this, we first should show that the family of bivariate (conditional) copula densities contained in a given
multivariate distribution forms a \textit{compact set} in the space of continuous functions on $[0,1]^2$. Then, it can be shown that the same
finite parameter family of copulae can be used to derive a given level of approximation to all conditional copulae simultaneously.
\par
Here, we develop the approximation method used by Bedford et al. (2012) to approximate any log-density function at
 the desired level of approximation which is more accurate and exhibits better properties. We first
 introduce some notations. The basic assumption is that all densities are continuous. We denote $\mathcal{C}(Z)$
  as the space of continuous real valued functions on a space $Z$, where $Z=[0,1]^r$ for some $r$, and the corresponding norm on $\mathcal{C}(Z)$ is
given by
\[
|| f_{1\dots r} || = \sup |f_{1\dots r}(x_{1}, \dots, x_{r})| .
\]
The set of all possible 2-dimensional (conditional) copulae is denoted by
$$
{\mathcal C}(f) = \{ c_{ij|i_1 \dots i_r}: 1 \le i,j,i_1, \dots, i_r \le n, i, j \neq i_1, \dots, i_r \}
$$
where $c_{ij|i_1 \dots i_r}$ is the copula of the conditional density of $X_i, X_j$ given $X_{i_1}, \dots, X_{i_r}$.
\par
The famous Arzela-Ascoli theorem can be used to check the compactness of the following function space,
$K \subset C([0,1]^2)$. This space is relatively compact if the functions in $K$ are equicontinuous and pointwise bounded.
\par
It can be shown that the following two spaces are relatively compact (Bedford et al. (2012), Theorem 3).
$$
{\mathcal M }(f) = \{ f_{i|i_1 \dots i_r}: 1 \le i,i_1, \dots, i_r \le n, i \neq i_1, \dots, i_r   \},
$$
and
$$
{\mathcal B}(f) = \{ f_{ij|i_1 \dots i_r}: 1 \le i,j,i_1, \dots, i_r \le n, i, j \neq i_1, \dots, i_r\}
$$
where $f_{i|i_1 \dots i_r}$ is the conditional density of $X_i$ given  $X_{i_1}, \dots, X_{i_r}$, and
$f_{ij|i_1 \dots i_r}$ is the conditional density of $X_i, X_j$ given $X_{i_1}, \dots, X_{i_r}$.
\par
It is then straightforward to show that the set ${\mathcal C }(f) \subset  C([0,1]^2)$ is relatively
compact. In addition, since all the functions in ${\mathcal C }(f) $ are  positive and uniformly bounded
away from 0, the set ${\mathcal {LNC} }(f) = \{ \ln(g):g \in {\mathcal C }(f) \} \subset  C([0,1]^2)$ is
also relatively compact (see Bedford et al. (2012) for details and proofs).
\par
As a result, the set $C([0,1]^2)$ can be considered as a vector space, and in this context a base is simply
a sequence of functions $h_{1},h_{2}, \dots \in C([0,1]^2)$ such that any function $g \in C([0,1]^2)$ can be
written as $g=\sum_{i=1}^\infty \lambda_{i} h_{i}$. In other words, it can be shown that given $\epsilon>0$,
there is a $k$ such that any member of $ {\mathcal {LNC} }(f)$ (or ${\mathcal C }(f))$ can be approximated
to within error $\epsilon>0$ by a linear combination of $h_1,h_2, \dots, h_k$. There are lots of possible bases,
for example, the following polynomial series
\[
u, v, uv, u^2, v^2, u^2v uv^2, \dots.
\]
which was mainly used in Bedford et al. (2012).
\par
In the next section, we will improve this density approximation based on the minimum information techniques considerably using the orthonormal polynomial series and Legender multiwavelets instead the ordinary polynomial series as the basis functions. We also exhibit other nice properties of these approximations.
\par
It should be noticed that the approximated copula density by the method described above might not be a copula density itself. Therefore,
the resulting approximation needs to be transformed in such a way to obtain a copula. This can be done by
weighting the approximated density. One of the most effective weighting
schemes is the $D_{1}AD_{2}$ algorithm mentioned in the previous section. If we have a continuous positive
real valued function $A(u,v)$ on $[0,1]^2$ then there are continuous positive functions $d_1(u)$ and $d_2(v)$,
such that $d_1.d_2.A$ is a copula density, that is, it has uniform marginal distributions. This density is called
$C$-\emph{Projection} of $A$ and denoted by $\mathcal{C}(A)$. Bedford et al (2012) present the following
lemma at which it allows us to control the error made when approximating a copula by another function.\\
\par
\begin{lemma} Let $g$ be a non-negative continuous copula density. Given $\epsilon>0$ there is a $\delta$ such that if $||g-f||<\delta$ then
$||g-C(f) ||<\epsilon$.
\end{lemma}

Note that these reweighting functions have the same differentiability properties as the function $f$ being reweighted. This can be seen from
the integral equation that they satisfy:
\[
d^{(1)}(u)=\frac{1}{\int d^{(2)}(v)f(u,v)dv}~~~and~~~d^{(2)}(v)=\frac{1}{\int d^{(1)}(u)f(u,v)du}.
\]
Eventually, the term given in (1) can be used to see that good approximation of each conditional copula would result in a
good approximation of the multivariate density of interest.

\section{Building approximations using minimally informative distributions}
In this section, we give practical guide to build a minimally -
informative vine structure to approximate any multivariate
distribution. In the previous section, we present a method proposed
by Bedford et al. (2012) that all conditional copulae can be
approximated using linear combinations of basis functions. In this
section, we are going to address the issue of how the appropriate
parameter values can be chosen. We also introduce a practical and efficient alternative based on using the minimum information criterion that
lies very close to the approach described above. In other words, given the basis functions  $\{1, h_{1},\ldots, h_{k}\}:[0,1]^{2}\to \mathbb{R}$, we seek values $\lambda_1, \dots,
\lambda_k$ so that $\exp(\sum_1^k \lambda_i h_i)$ is close to the approximated copula density. This can be done by fitting the moments of $h_{i}$ in the minimum information framework. Therefore, if $E_{g}[h_{i}(u,v)] = \alpha_i$, we seek for the minimum information copula density that also has these moments. This copula density can uniquely be determined, using the $D_{1}AD_{2}$ algorithm, as follows
$$
d^1(u)d^2(v) \exp( \sum_1^k \lambda_i h_i(u,v)).
$$
As mentioned above, a multivariate distribution can be modelled by a vine structure where it can be defined as a
 decomposition of the given multivariate distribution into certain conditional copulae, associated with the conditioned
  and conditioning sets of the vine. The following algorithm is summarised the steps to approximate the given multivariate
   distribution associated with a vine structure:

\begin{enumerate}
\item Specify a basis family, denoted by $\mathcal{S}(k)=\{h_{1},h_{2},\ldots\}$
\item Specify a vine structure
\item For each part of vine, the bivariate copulae, specify either
\begin{itemize}
\item mean $\alpha_{1},\ldots,\alpha_{k}$ for $h_{1},\ldots,h_{k}$ on each pairwise copula;
\item functions $\alpha_{m}(ji\mid D_{e})$  for the mean values as functions of the conditioning variables, for $m=1,\ldots,k$.
\end{itemize}
\end{enumerate}
\par
One of the main aspect that would effect the aforementioned approximation
is the basis family. Here, we examine the impact of two basis
families, the \textit{orthonormal polynomial} series and \textit{Legender
multiwavelets} on approximating the minimum information copulae
and the multivariate distribution associated with the chosen vine structure. We first briefly introduce these two basis
functions.
%

\subsection{Constructing Orthonormal Polynomial base}\label{basisfunctions}

\par
In mathematics, particularly numerical analysis, a basis function is
an element of the basis for a function space. The term is a
degeneration of the term basis vector for a more general vector
space; that is, each function in the function space can be
represented as a linear combination of the basis functions. We say
two polynomial functions $g_{1}$ and $g_{2}$ are orthonormal
polynomial in the interval $[0,1]$, if
\begin{equation}
\int_{0}^{1}g_{1}(u)g_{2}(u)du=\left\{ \begin{array}{ll}
1 & ~~~\mbox{for~~ $g_{1}(u)=g_{2}(u)$};\\
0 & ~~~\mbox{for~~ $g_{1}(u)\neq g_{2}(u)$}.
\end{array} \right.\label{01Loss}
\end{equation}
Orthonormal polynomial base can be more convenient than some natural
basis for the purpose of calculation. In fact, if the basis is an
orthonormal polynomial basis, adding a new item to the expansion
does not change coefficient of the already found shorter expansion
(Gui, 2009). But if the basis is not orthonormal, any new item has
in general nonzero projection on previous items. It means that the
already found coefficients of the expansion would have to be
changed. That is one of the reason we use orthonormal polynomial
basis functions as the basis family, $\mathcal{S}(k)$. It is reasonable to consider Gram-Schmidt orthonormal polynomial
basis which is one of the famous orthonormal polynomial
basis functions on $[0,1]$.
\par
To construct this orthonormal polynomial basis over the interval $[0,1]$, we use the \textit{Gram-Schmidt process} as follows.
\begin{eqnarray*}
\varphi_{0} (u)=1,\\
\varphi_{n} (u)=\frac{u^{n}-\sum_{j=0}^{n-1}\frac{\int_{0}^{1}u^{n} \varphi_{j} (u)du}{\int_{0}^{1} \varphi_{j}^{2} (u)du}\varphi_{j}(u)}{||u^{n}-\sum_{j=0}^{n-1}\frac{\int_{0}^{1}u^{n} \varphi_{j} (u)du}{\int_{0}^{1} \varphi_{j}^{2} (u)du}\varphi_{j}(u)||}~~~~n\geq 1
\end{eqnarray*}
The first few functions are
\begin{eqnarray*}
\varphi_{0} (u)=1,~~~\varphi_{1}(u)=\sqrt{3} (-1+u),~~~\varphi_{2} (u)=\sqrt{5} (1-6u+6u^2),\\
\varphi_3 (u)=\sqrt{7} (-1+12u-30u^2+20u^3 ),~~~~\varphi_4 (u)=\sqrt{9} (1-20u+90u^2-140u^3+70u^4 )\\
\varphi_5 (u)=\sqrt{11} (-1+30u-210u^2+560u^3-630u^4+252u^5),~~~\ldots
\end{eqnarray*}

\subsection{Constructing Legender Multiwavelets base}\label{basisfunctions}

The use of wavelets for density estimation has recently gained in popularity due to their ability to approximate a large class of functions,
including those with localized, abrupt variations. However, a well-known attribute of wavelet bases is that they can not be simultaneously symmetric, orthogonal, and compactly supported. Therefore, a more general, vector-valued, construction of
wavelets is proposed by Locke and Peter (2012) to overcome this disadvantage, and making them natural choices for estimating
density functions, many of which exhibit local symmetries around features such as a mode. Locke and Peter (2012) introduce the methodology of wavelet density estimation using multiwavelet bases and illustrate several empirical results where multiwavelet estimators outperform
their wavelet counterparts at coarser resolution levels. 
\par
In this section, we use the multiwavelet bases to approximate the minimum information copula. The main advantage of using these bases over the polynomial bases introduced in the previous subsection is that the wavelets (and in particular, multiwavelets) are are better choices where the functions of interest contain discontinuities and sharp spikes. In addition, in order to preserve the orthonormality property among the multiwavelet bases, we use Legender multiwavelet bases. 
\par
In order to construct these bases, we need to introduce some notions and definitions which are briefly described in the following subsections.
\subsubsection{Multiresolution analysis}\label{Legender multiwavelets}
Wavelet theory is based on the idea of multiresolution analysis (MRA). Usually
it is assumed that an MRA is generated by one scaling function, and dilates
and translates of only one wavelet $\phi\in L^{2}(\mathbb{R})$ form a stable basis of $L^{2}(\mathbb{R})$. 
\par
We can generate a reference subspace or sample space $V_{0}$ as $L^{2}$-closure of the linear span of the integer translation of the following functions
$\phi^{m}\in L^{2}(R), m=0,\ldots,r$, namely
\[
V_{0}=clos_{L^{2}}\prec\phi^{m}(.-k):k\in Z\succ,~~~~~~~m=0,\ldots,r,
\]
and consider subspace
\[
V_{j}=clos_{L^{2}}\prec\phi_{j,k}^{m}:k\in Z\succ,~~~~~~~j \in Z~\textrm{and}~~ m=0,\ldots,r,\\
\]
where
$\phi_{j,k}^{m}=\phi^{m}(2^{j}x-k):j,k\in Z,~~~~~~~m=0,\ldots,r.$
\par
Now, we are able to present a proper definition of multiresolution analysis as follows.
\par
\textbf{Definition~1:}~~Functions $\phi^{m}\in L^{2}(R)$,
are said to generate a multiresolution analysis (MRA) if they
generate a nested sequence of closed subspaces $V_{j}$ that
satisfies
\begin{equation}
\begin{array}{lcr}
i) & ~~~\mbox{$...\subset V_{-1}\subset V_{0}\subset V_{1}\subset...$}\\
ii) & ~~~\mbox{$clos_{L^{2}}(\bigcup_{j\in Z}V_{j})=L^{2}(R) $} \\
iii) &~~~\mbox{$\bigcap_{j\in Z}V_{j}=0$}\\
iv) & ~~~\mbox{ $\phi^{m}(x)\in V_{j} \Longleftrightarrow \phi^{m}(x+2^{-j})\in V_{j} \Longleftrightarrow \phi^{m}(2x)\in V_{j+1}$}\\
v) & ~~~\mbox{ $\{\phi^{m}(.-k)\}_{k\in Z}; $ ~~~~~~ \textrm{form a Riesz basis of} $V_{0}$}\\
\end{array} .\label{01Loss}
\end{equation}
If $\phi^{m}$ generates an MRA, then $\phi^{m}$ are called \emph{scaling
functions}. In case that the different integer translate of
$\phi^{m}$ are orthogonal (with respect to the standard linear
product $\prec f, g \succ
=\int_{-\infty}^{+\infty}{f(x)\overline{g(x)}dx}$) for two functions
in $L^{2}(R)$, denoted by $\phi^{m}(.-k)\bot \phi^{\overline{m}} (.-\overline{k})$
for $m\neq \overline{m},~k\neq \overline{k},$ the scaling functions are called
an orthogonal scaling functions.
\par
As the subspaces $V_{j}$ are nested, there exist complementary
orthogonal subspaces $W_{j}$ such that
$$
V_{j+1}=V_{j}\bigoplus W_{j}, ~~~~ j\in Z
$$
here and in the following $\bigoplus$ denotes orthogonal sums.
\par
This yields an orthogonal decomposition of $L^{2}(R)$, namely;
$$
L^{2}(R) = \bigoplus_{j \in Z} W_{j},
$$
\par
\textbf{Definition~2:}~~Functions $\psi^{m}\in L^{2}(R)$ are
called wavelets, if they generate the complementary orthogonal
subspaces $W_{j}$ of a MRA, i.e.,
\[
W_{j}=clos_{L^{2}}\prec\psi_{jk}^{m}:k\in~Z\succ,~~~~~~~j\in Z, \textrm{and}~~ m=0,\ldots,r,
\]
where $\psi_{j,k}^{m}=\psi^{m}(2^{j}x-k), j,k \in Z.$
\par
Obviously, $\psi_{j,k}^{m}\perp
\psi_{\overline{j},\overline{k}}^{\overline{m}}$ for $j\neq
\overline{j}$, $m\neq \overline{m}$ and $k\neq \overline{k}$, if
$\prec 2^{j/2}\psi _{j,k}^{m},~2^{\overline{j}/2} \psi
_{\overline{j},\overline{k}}^{\overline{m}}\succ=\delta_{j,\overline{j}}\delta_{k,\overline{k}}\delta_{m,\overline{m}}$,
then $\psi^{m}$ are called \emph{orthogonal wavelets}, where\\
$$
\delta_{i,k}=\left\{ \begin{array}{ll}
1 & ~~~\mbox{for~~ $i=k$};\\
0 & ~~~\mbox{for~~ $i\neq k$}.
\end{array} \right.\label{01Loss}
$$
Now, we are able to define Legender scaling functions and its corresponding
multiwavelets according to MRA definition give above.
\subsubsection{Construction of Scaling  Functions}\label{MRA}
Legendre multiwavelets system with multiplicity $r$ consists of $r$
scaling functions and $r$ wavelets. The $r$-th order Legendre
scaling functions are the set of $r+1$ functions
$\phi^{0}(x),\ldots,\phi^{r}(x)$ where $\phi^{i}(x)$ is a polynomial of
$i$-th order and all $\phi$'s form orthogonal basis (Shamsi and Razzaghi, 2005), that is,
for $i=0,1,\ldots,r,$
\begin{equation}
\phi^{i}(x)=\sum_{k=0}^{i}a_{ik}x^{k}, ~~~~ \textrm{for}~~i=0,1,\ldots,r
\end{equation}
The coefficient $a_{ik}$ are chosen so that $a_{ik}\geq 0$, and
\begin{equation}
\int_{0}^{1}\phi^{i}(x) \phi^{k}(x)dx =\delta_{i,k}, ~~~~ \textrm{for}~~
i,k=0,1,...,r
\end{equation}
The scaling functions $\phi^{i}(x)$ have symmetry, anti-symmetry
properties for odd or even $i$, respectively. The two-scale
relations for Legendre scaling functions of order $r$, are in the
form (Albert et al., 2002);
\begin{equation}
\phi^{i}(x)=\sum_{j=0}^{r}p_{i,j}\phi^{j}(2x)+\sum_{j=0}^{r}p_{i,r+j+1}\phi^{j}(2x-1),
~~~~\textrm{for}~ i=0,1,\ldots,r
\end{equation}
The coefficients $p_{i,j}$ determined uniquely by substituting
equation (8) to (10). Now we would like to mention two remarks on
the two scale relations.
\begin{enumerate}
\item Since $\phi^{i}(x)$ is a $i$-th order polynomial, the right hand
side of (10) has at most $i$-th order scaling functions. Therefore,
$p_{i,j}=p_{i,r+j+1}=0$ for $i<j$.
\item The two scale relations for the Legendre scaling function of
order $n$ which is lower than r is a subset of first $n$ two-scale
relations for $\phi^{i}$ for $i=0,1,\ldots,n$ form $r$-th order two
scale relations.
\end{enumerate}
\subsubsection{Construction of Wavelets}\label{MRA}
\par
The two-scale relation for the $r$-th order Legendre multiwavelets is given in the following form (Albert et al., 2002):
\begin{equation}
\psi^{i}(x)=\sum_{j=0}^{r}q_{i,j}\phi^{j}(2x)+\sum_{j=0}^{r}q_{i,r+j+1}\phi^{j}(2x-1),
~~~~ \textrm{for}~~ i=0,1,\ldots,r.\label{two-scale}
\end{equation}
The $2(r+1)^{2}$ unknown coefficients $\{q_{i,j}\}$ in (\ref{two-scale}) can be determined in terms of the following $2r(r+1)$ vanishing moment conditions (\ref{VM}) and $2(r+1)$ orthongonal conditions (\ref{OR}).
\par\begin{description}
      \item[Vanishing moments]
\begin{equation}
\int_{0}^{1}\psi^{i}(x)x^{j}dx=0, ~~~~ \textrm{for}~~ i=0,1,\ldots,r;
~j=0,1,...,i+r.\label{VM}
\end{equation}
 \item[Orthogonality]
\begin{equation}
\int_{0}^{1}\psi^{i}(x)\psi^{j}(x)dx=\delta_{i,j}, ~~~~ \textrm{for}~~
i,j=0,1,\ldots,r.\label{OR}
\end{equation}
\end{description}
For example, the Legendre scaling functions of order 5 consist of 6
functions as follows:
\begin{equation}
\begin{array}{lcr}
\phi^{0}(x)=1 & ~\mbox{for~$0\leq x \leq 1$}\\
\phi^{1}(x)=\sqrt{3}(-1+2x) & ~\mbox{for~$0\leq x \leq 1$}\\
\phi^{2}(x)=\sqrt{5}(1-6x+6x^{2}); & ~\mbox{for~$0\leq x \leq 1$} \\
\phi^{3}(x)=\sqrt{7}(-1+12x-30x^{2}+20x^{3}) &~\mbox{for~$0\leq x \leq 1$}\\
\phi^{4}(x)=\sqrt{9}(1-20x+90x^{2}-140x^{3}+70x^{4}) & ~\mbox{ for~$0\leq x \leq 1$}\\
\phi^{5}(x)=\sqrt{11}(-1+30x-210x^{2}+560x^{3}-630x^{4}+252x^{5}) & ~\mbox{ for~$0\leq x \leq 1$}\\
\end{array} \label{01Loss}
\end{equation}
\par
The closed form solution to the Legendre multiwavelets of order 5,
$\psi^{0}(x),\psi^{1}(x) \psi^{2}(x),\psi^{3}(x), \psi^{4}(x)$ and
$\psi^{5}(x)$ are given below which are determined using the
conditions (\ref{VM}) and (\ref{OR}).
\[
\psi^{0}(x)=\left\{ \begin{array}{ll}
3.55-146.72x+1419.86x^{2}-5300.81x^{3}+8519.15x^{4}-4997.9x^{5} & ~\mbox{for~$0\leq x \leq  \frac{1}2 $};\\
 -502.87+4122.32x-13346.68x^{2}+21203.23x^{3}-16470.37x^{4}+4997.907x^{5} & ~\mbox{for~$ \frac{1}2 \leq x \leq 1$}
\end{array} \right.\label{02Loss}
\]
\[
\psi^{1}(x)=\left\{ \begin{array}{ll}
-3.47+181.55x-2188.78x^{2}+10023.38x^{3}-19433.09x^{4}+13500.89x^{5} & ~\mbox{for~$0\leq x \leq \frac{1}2 $}\\
 -2080.47+15646.19x-46291.67x^{2}+67299.87x^{3}-48071.33x^{4}+13500.89x^{5} & ~\mbox{for~$\frac{1}2 \leq x \leq 1$} 
\end{array} \right.\label{03Loss}
\]
\[
\psi^{2}(x)=\left\{ \begin{array}{ll}
 2.81-174.03x+2438.52x^{2}-12760.78x^{3}+27823.96x^{4}-21415.36x^{5}& ~\mbox{for~$0\leq x \leq \frac{1}2 $}\\
 -4084.87+29360.26x-83053.61x^{2}+1.16\times10^5x^{3}-79252.82x^{4}+21415.36x^{5}& ~\mbox{for~$\frac{1}2 \leq x \leq 1$}
\end{array} \right.\label{04Loss}
\]
\[
\psi^{3}(x)=\left\{ \begin{array}{ll}
1.71-121.14x+1911.69x^{2}-11113.58x^{3}+26588.59x^{4}-22203.27x^{5}  & ~\mbox{for~$0\leq x \leq \frac{1}2 $}\\
4935.99-34300.49x+93930.24x^{2}-1.27\times10^5x^{3}+84427.78x^{4}-22203.27x^{5} & ~\mbox{for~$\frac{1}2 \leq x \leq 1$}
\end{array} \right.\label{05Loss}
\]
\[
\psi^{4}(x)=\left\{ \begin{array}{ll}
-0.71+56.63x-998.10x^{2}+6413.33x^{3}-16797.83x^{4}+15222.11x^{5}  & ~\mbox{for~$0\leq x \leq \frac{1}2$}\\
3895.43-26219.63x+69675.97x^{2}-91443.07x^{3}+59312.70x^{4}-15222.11x^{5} & ~\mbox{for~$\frac{1}2 \leq x \leq 1$}
\end{array} \right.\label{06Loss}
\]
\[
\psi^{5}(x)=\left\{ \begin{array}{ll}
 0.17-15.67x+308.12x^{2}-2193.38x^{3}+6324.24x^{4}-6273.06x^{5} & ~\mbox{for~$0\leq x \leq \frac{1}2$}\\
 1849.58-12047.91x+31057.19x^{2}-39627.04x^{3}+25041.07x^{4}-6273.06x^{5} & ~\mbox{for~$\frac{1}2 \leq x \leq 1$}
\end{array} \right.\label{07Loss}
\]
\section{Application: Norwegian Financial returns}
In this section, we apply the approximation method presented in this paper using the basis functions introduced in the previous section as the basis families, $\mathcal{S}(k)$ (as mentioned in the first step in the algorithm above) to approximate the multivariate
distribution associated with the selected vine structure corresponding to the Norwegian financial returns. We then exhibit the potential flexibility of our approach by comparing it with the other methods cited in Bedford et al. (2012) and Aas et al. (2009).\\
\par
\textbf{Example:}~~In this example we use the same data set as considered by Aas et al. (2009) and Bedford et al. (2012) to
illustrate the approximation method introduced in this paper. The
data consists of four time series of daily data: the Norwegian stock
index (TOTX), the MSCI world stock index, the Norwegian bond index
(BRIX) and the SSBWG hedged bond index. They are recorded over the
period 04.01.1999 to 08.07.2003 at which 1094 data are collected. We
denote these four variables $T,B,M$ and $S$, respectively.

\begin{figure}
\begin{center}
\vspace{-6cm} \scalebox{0.65}{\includegraphics{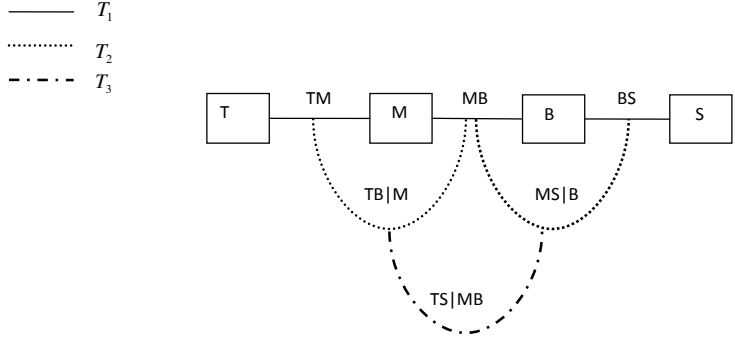}}
\end{center}
\vspace{-6.2cm} \caption{Selected vine structure for the Norwegian
stock data set with 4 variables: Norwegian stock index (T), MSCI
world stock index (M), Norwegian bond index (B) and SSBWG hedged
bond index (S)} \label{vine4NFM}
\end{figure}
\par
We first shall remove serial correlation in these four time series,
that is, the observation of each variable must be independent over
time. Hence, the serial correlation in the conditional mean and the
conditional variance are modeled by an AR(1) and a GARCH(1,1) model
(Bollerslev, 1986), respectively. Thus, the following model for log-return $x_i$ is considered for the $i^{th}$ time series
\begin{eqnarray*}
x_{i,t}=c_{i}+\alpha_{i} x_{i,t-1}+\sigma_{i,t} z_{i,t}\\
E[z_{i,t} ]=0~~~\textrm{and}~~~ Var[z_{i,t}]=1\\
\sigma^{2}_{i,t}=\alpha_{i,0}+a_{i}\epsilon_{i,t-1}^{2}+b{_i} a\sigma_{i,t-1}^2
\end{eqnarray*}
where $\epsilon_{i,t-1}=\sigma_{i,t}+z_{i,t}$ (see Aas et al., 2009).
\par
The further analysis is performed on the standardized residuals
$z_i$ . If the AR(1)-GARCH(1,1) models are successful at modeling
the serial correlation in the conditional mean and the conditional
variance, there should be no autocorrelation left in the
standardized residuals and squared standardized residuals. We can use the modified Q-statistic
and the Lagrange multiplier test, respectively, to check this (Aas et al, 2009).
For all series, the null hypothesis that there is no autocorrelation
left for the both tests cannot be rejected at the \%5 level. Since, we are mainly interested in estimating the dependence structure of
the risk factor, the standardized residual vectors are converted to
the uniform variables using the kernel method before further
modeling. We denote the converted time series of $T,M,B$ and $S$ by
$X,Y,Z$ and $U$, respectively.\\
\par
Here, we are going to derive the vine approximation fitted to this
data set to any given multivariate density using minimum information
distribution. We adopt a vine structure to these data, as presented
in Figure \ref{vine4NFM}. Note that, the
corresponding functions of the copula variables $X$, $Y$, $Z$ and
$U$ associated with $T, M, B, S$ can be derived. For instance, these are defined by
$h_{i}(X,Y)=h'_{i}(F_{1}^{-1}(X), F_{2}^{-1}(Y))$ and should also have the same specified expectation, that is,
$E(h'_{i}(T,M))= E(h_{i}(X,Y))$. We derive the minimum information
copulae calculated in this example based on them the copula variables, $X,Y, X, W$. We initially construct minimally informative copulas between each
set of two adjacent variables in the first tree, $T_1$.  Therefore, it is essential to decide which bases should be taken and how many discretization
points should be used in each case. We start illustrate our procedure for the first copula in the first tree between $T, M$.

\begin{figure}[h]
\begin{center}
\scalebox{0.5}{\includegraphics{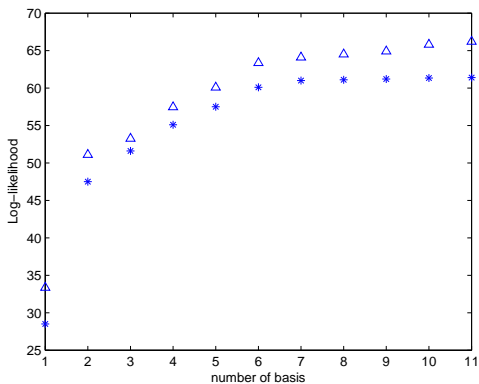}}
\end{center}
\caption{The log-likelihood of the minimally informative copula between $T$ and $M$, calculated based on orthonormal basis (*) and Legendre multiwavelets ($\triangle$).\label{loglikelihood}}
\end{figure}
\par
We could simply choose basis functions, starting with simple
orthonormal polynomials or Legendre multiwavelets basis, and moving
to more complex ones, and include them until we are satisfied with
our approximation. We included the following orthonormal polynomial
basis functions, constructed using Gram-Schmidt process, in
order
\begin{eqnarray*}
\varphi_1 (x) \varphi_1 (y),\varphi_1 (x) \varphi_2 (y),\varphi_2 (x) \varphi_1 (y),\varphi_1 (x) \varphi_3 (y),\varphi_3 (x) \varphi_1 (y),\\
\varphi_2 (x) \varphi_2 (y),\varphi_2 (x) \varphi_3 (y),
\varphi_3 (x) \varphi_2 (y),\varphi_1 (x) \varphi_4 (y),\varphi_4 (x) \varphi_1 (y),\\
\varphi_1 (x) \varphi_5 (y),\varphi_5 (x) \varphi_1 (y), \varphi_2
(x) \varphi_4 (y),\varphi_4 (x) \varphi_2 (y),\varphi_3 (x)
\varphi_3 (y),\ldots
\end{eqnarray*}
and the following Legendre multiwavelets basis functions which is
constructed based on the method presented in subsection 5.2
\begin{eqnarray*}
\phi^{1}(x)\psi^{0}(y),\phi^{1}(x)\psi^{1}(y),\phi^{2}(x)\psi^{0}(y),\phi^{1}(x)\psi^{2}(y),\phi^{2}(x)\psi^{1}(y),\\
\phi^{3}(x)\psi^{0}(y),\phi^{1}(x)\phi^{1}(y),\psi^{1}(x)\psi^{1}(y),\psi^{0}(x)\psi^{2}(y),\psi^{2}(x)\psi^{0}(y),\\
\phi^{1}(x)\phi^{2}(y),\phi^{2}(x)\phi^{1}(y),\varphi^{0}(x)\varphi^{3}(y),\varphi^{3}(x)\varphi^{0}(y)
\varphi^{1}(x)\varphi^{2}(y),\ldots
\end{eqnarray*}
Bedford et al. (2012) show that adding the basis functions in this
way is not optimal, and propose a method which is similar to a
stepwise regression. In this method, at each stage, we propose to
assess the log-likelihood of adding each additional basis function.
We then include the function which produces the largest increase in
the log-likelihood. At moment, we are investigating some other
methods, such as, Genetic, PSO algorithm, Lasso and ant-colony
algorithms, to find the most optimal basis functions in a sense that
with smaller number of these bases, we would get the largest
log-likelihood.
\par
Figure \ref{loglikelihood} shows the changes of log-likelihood in terms of adding basis functions for
orthonormal polynomial ($*$) and Legendre multiwavelets ($\triangle$). In order to compare our results with the
approximations made in Bedford et al. (2012) using the ordinary
polynomial series, we choose six orthonormal basis functions using
the stepwise method as follows
\[
\varphi_1 (T) \varphi_1 (M), \varphi_2 (T) \varphi_2 (M), \varphi_1
(T) \varphi_2 (M), \varphi_1 (T) \varphi_3 (M), \varphi_3 (T)
\varphi_3 (M), \varphi_4 (T) \varphi_1 (M)
\]
and also we choose six Legendre multiwavelet basis functions as follows
\[
\phi ^{1}(T)  \phi^{1}(M), \phi ^{2}(T) \phi ^{2}(M), \phi^{4} (T)
\phi^{5} (M), \phi ^{1}(T) \phi^{2}(M), \phi^{3}(T)\phi ^{3}(M),
\psi^{2} (T) \phi^{4}(M)
\]
The corresponding log-likelihood based on orthonormal plynomial functions reaches to 60.66 and based on Legendre multiwaveletswhich reaches to
63.36 which both are more than the log-likelihood, 58.1256, based on six basis functions calculated in Bedford et al. (2012). The corresponding expectations of the selected orthonormal plynomial basis functions using the Norwegian financial returns data are calculated as
\begin{eqnarray*}
\alpha_{1}=-0.2292,~~ \alpha_{2}=0.2104,~~\alpha_{3}=0.0808,~~\alpha_{4}=-0.1025,~~\alpha_{5}=-0.1120,~~\alpha_{6}=0.0463
\end{eqnarray*}
and also for the selected Legendre multiwavelet bases are given by
\begin{eqnarray*}
\alpha_{1}=0.4803,~~\alpha_{2}=0.2298,~~\alpha_{3}=-0.0021,~~\alpha_{4}=0.0194,~~\alpha_{5}=0.0866,~~\alpha_{6}=0.0191,
\end{eqnarray*}
We now able to construct the minimum information copula $C_{TM}$ with respect to the uniform distributions given the constraints as the selected basis functions reported above by the method described in this paper. We first need to determine the number of
discretization points (grid size). It is trivial to conclude that a
larger grid size will provide a better approximation to the
continuous copula but at the cost of more computation time.
Similarly, the approximation will become more precise if we run the
$D_{1} AD_{2}$ algorithm in more iterations. Indeed, this would cost us more
computation time. Bedford et al. (2012) show that the number of
iterations needed will also depend on the grid size. The considered
errors are reported to be in the range $1\times 10^{-1}$ to $1\times
10^{-24}$. Thus, the larger the number of grid points used, the
larger the number of iterations that are needed for convergence
which is true over all error levels. The grid sizes all follow the
same pattern with large increases in the number of iterations needed
for improved accuracy initially and smaller increases when the error
is smaller. We choose a grid size of $200\times200$ throughout of
this example.
\par
\begin{figure}[ht]
\begin{minipage}[b]{0.45\linewidth}
\centering
\includegraphics[width=\textwidth]{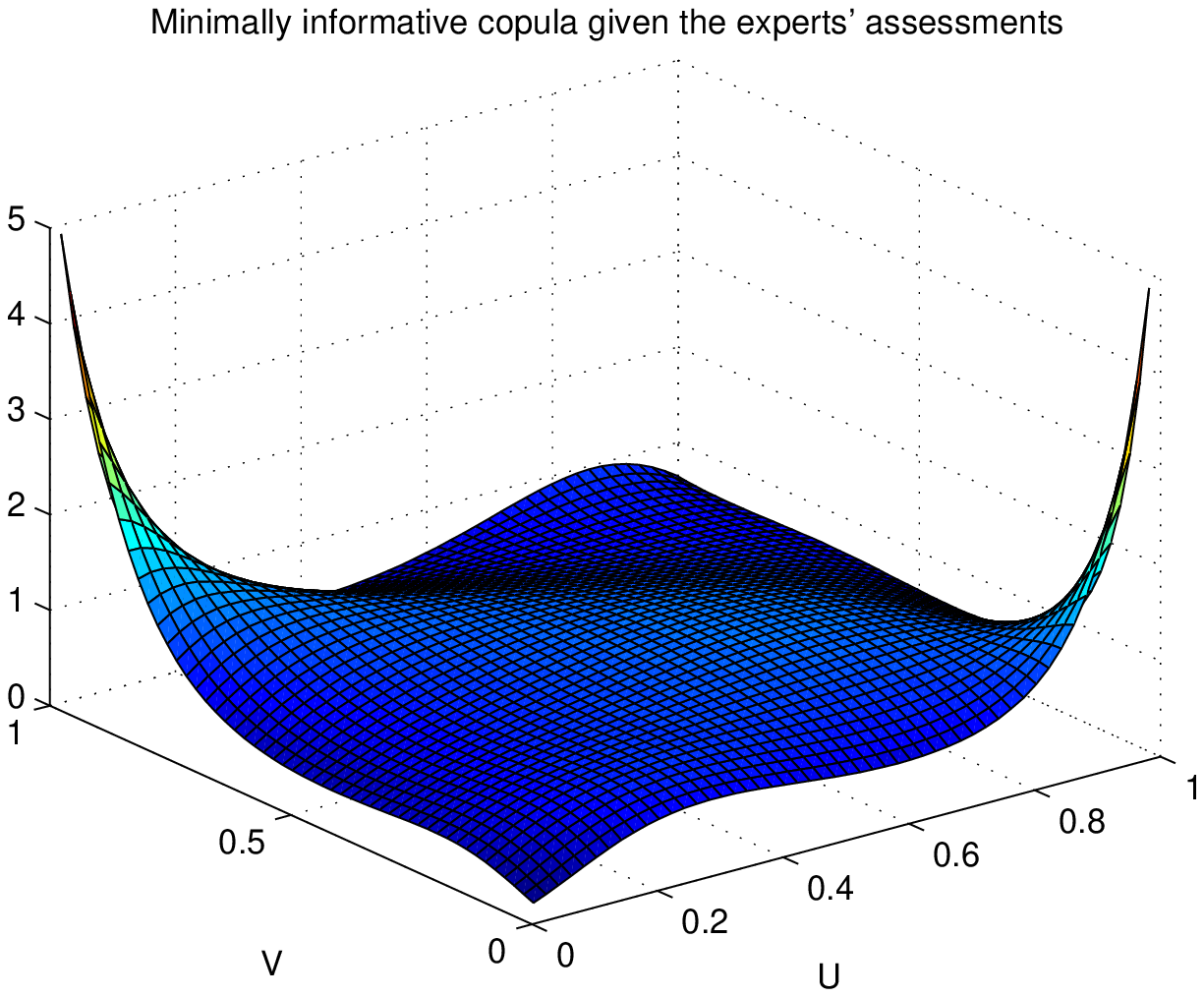}
\caption{The minimally informative copula between $T$ and $M$ using the orthonormal polynomial bases}
\label{T-Morthogonal}
\end{minipage}
\hspace{0.5cm}
\begin{minipage}[b]{0.45\linewidth}
\centering
\includegraphics[width=\textwidth]{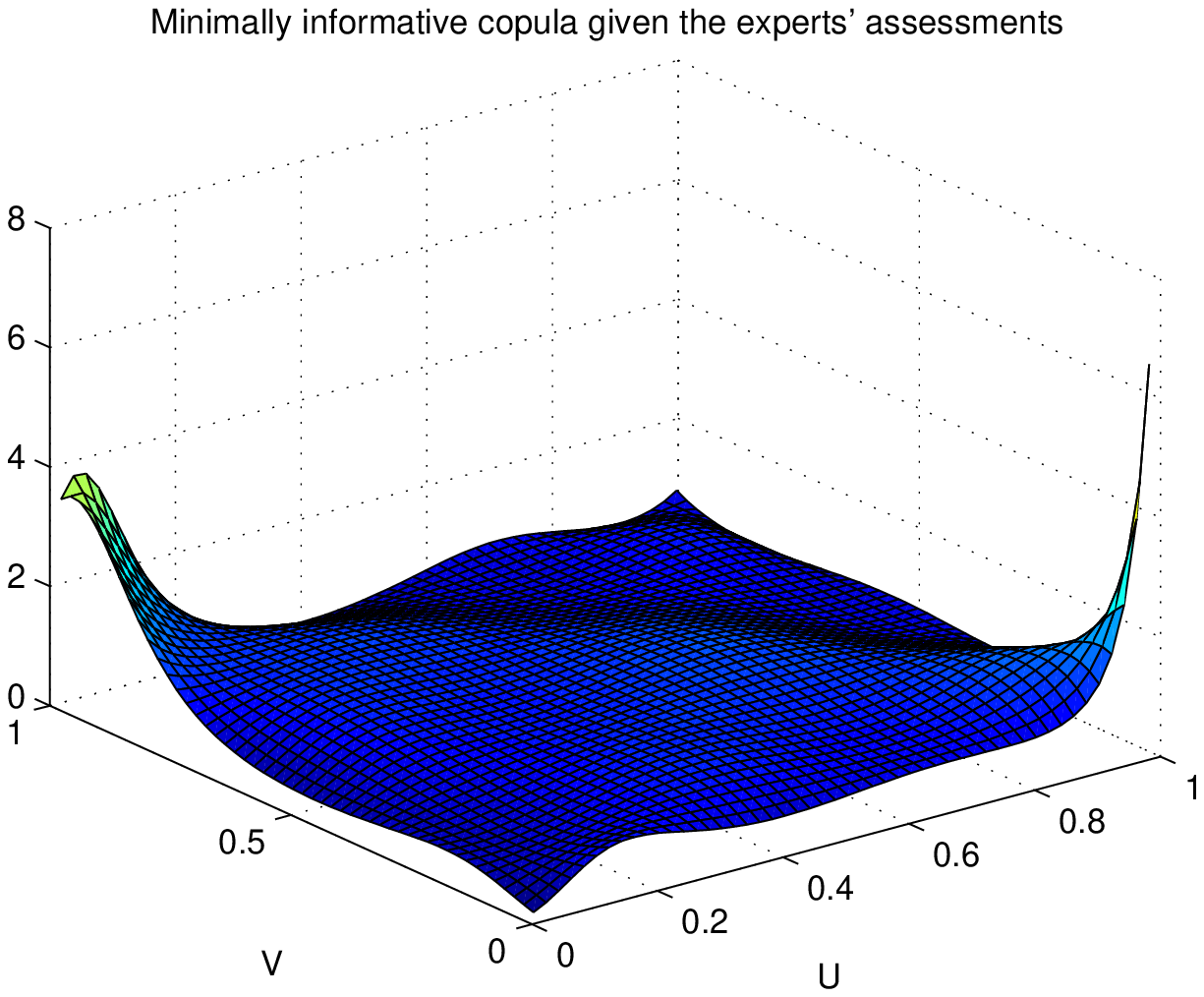}
\caption{The minimally informative copula between $T$ and $M$ using the Legendre multiwavelets bases}
\label{T-Mwavelet}
\end{minipage}
\end{figure}

Based on the information given above regarding the grid size, number
of iterations and error size, we can derive the minimum information
copula $C_{TM}$ associated with the chosen constraints. This copula
based on the orthonormal polynomial bases is plotted in Figure \ref{T-Morthogonal}, and the copula based on the
Legendre multiwavelet basis functions is plotted in Figure \ref{T-Mwavelet}. We present Lagrange
multiplies values (or parameter values) for this approximated copula density as follows
\[
\lambda_{1}=-0.1995, \lambda_{2}=0.1651 ,\lambda_{3}=0.0912, \lambda_{4}=-0.0774 ,\lambda_{5}=-0.0772 ,\lambda_{6}=0.0527
\]
and in the similar way these parameter values for the minimum information copula based on the Legendre multiwavelets bases are given by
\[
\lambda_{1}=1.9845, \lambda_{2}=1.6158 ,\lambda_{3}=0.0023, \lambda_{4}=-0.0263 ,\lambda_{5}=-7.4167 ,\lambda_{6}=3.6819
\]

One of the main advantages of using the orthonormal polynomial and Legendre multiwavelets basis functions
over the ordinary polynomial series considered in Bedford et al.
(2012) is that the $D_{1}AD_{2}$ algorithm converges faster using
these bases. This is because of the nice property of these two bases that adding
a new basis to the kernel defined in (3) and used to construct the minimum
information copula, does not change the Lagrange multipliers
of the already used in the kernel. This is shown in Table \ref{Table1} for the orthonorma polynomial basis functions. But, this is not the case when
one is applying the ordinary polynomial bases (as proposed by Bedford et al,
2012) to calculate the minimum information copula. In this situation,  we need to run the $D_{1}AD_{2}$ algorithm each time  a new
base is added to the already chosen bases, and the parameter values are changing accordingly. Therefore, more
iterations are required for the $D_{1}AD_{2}$ algorithm to converge. The optimisation time required for the $D_{1}AD_{2}$
algorithm using the the orthonormal polynomial bases is only 35.8646 seconds, for Legendre multiwavelets is 29.359
while this time for the ordinary polynomial bases is 72.93 seconds which is almost twofold of the former one and almost two and half times more than the latter one.
\par
\begin{table}
\begin{center}
\begin{tabular}{|l|l|l|}
\hline
\multicolumn{1}{|c|}{Base}& \multicolumn{1}{|c|}{Parameter values} & \multicolumn{1}{|c|}{Log-Likelihood}\\
\hline
$\varphi_{1} (T) \varphi_{1} (M)$ & -0.1995 & 29.36 \\
Previous one, $\varphi_{2} (T) \varphi_{2} (M)$ & -0.1995,   0.1651 & 49.2\\
Previous one, $\varphi_{1} (T) \varphi_{2} (M)$ & -0.1995, 0.1651,  0.0912 &52.8\\
Previous basis,  $\varphi_{1} (T) \varphi_{3} (M)$ &    -0.1995,   0.1651,   0.0912,   -0.0774&     56.16\\
Previous basis, $\varphi_{3} (T) \varphi_{3} (M)$ &-0.1995,   0.1651,   0.0912,   -0.0774,   -0.0772& 59.04\\
Previous basis, $\varphi_{4} (T) \varphi_{1} (M)$ & -0.1995,   0.1651,   0.0912,   -0.0774,   -0.0772,   0.0527 & 60.66\\ \hline
 \end{tabular}
\caption{Adding new orthonormal polynomial basis did not change Lagrange multiplier}\label{Table1}
\end{center}
\end{table}
Furthermore, by comparing the log-likelihoods of the minimum information copulas based on the ordinary polynomial, orthonormal polynomial and Legangre multiwavelets, we can conclude that the latter one produce more reliable copula density approximation in the sense that the corresponding log-likelihood
is much larger. We present the log-likelihood of these approximated copulae using the aforementioned bases in Table \ref{Table0}.
\begin{table}
\begin{center}
\begin{tabular}{|l|l|l|}
\hline
\multicolumn{1}{|c|}{Type of Bases}& \multicolumn{1}{|c|}{Number of bases} & \multicolumn{1}{|c|}{Log-Likelihood}\\
\hline
Ordinary Polynomial (Bedford et al. 2012) & 6 & 58.1256\\
Orthonormal polynomial (Subsection 5.1) & 6 & 60.66\\
Legangre multiwavelets (Subsection 5.2) &  6&     63.36\\
 \hline
 \end{tabular}
\caption{Log-likelihoods of the minimum information copulae of different basis functions}\label{Table0}
\end{center}
\end{table}
It should be noticed that the log-likelihood of the approximated copula using only 5 bases of orthonormal polynomial or Legangre multiwavelets is still larger than the fitted copula based on the six ordinary polynomial
bases. In addition, we realize that the derived
approximated copula in term of the bases proposed in this paper are more
flexible than ordinary polynomial bases, since they aren't sensitive to the initial values chosen for the parameter values (Lagrange multipliers)
in the $D_{1}AD_{2}$ algorithm.
\par
\begin{figure}[ht]
\begin{minipage}[b]{0.45\linewidth}
\centering
\includegraphics[width=\textwidth]{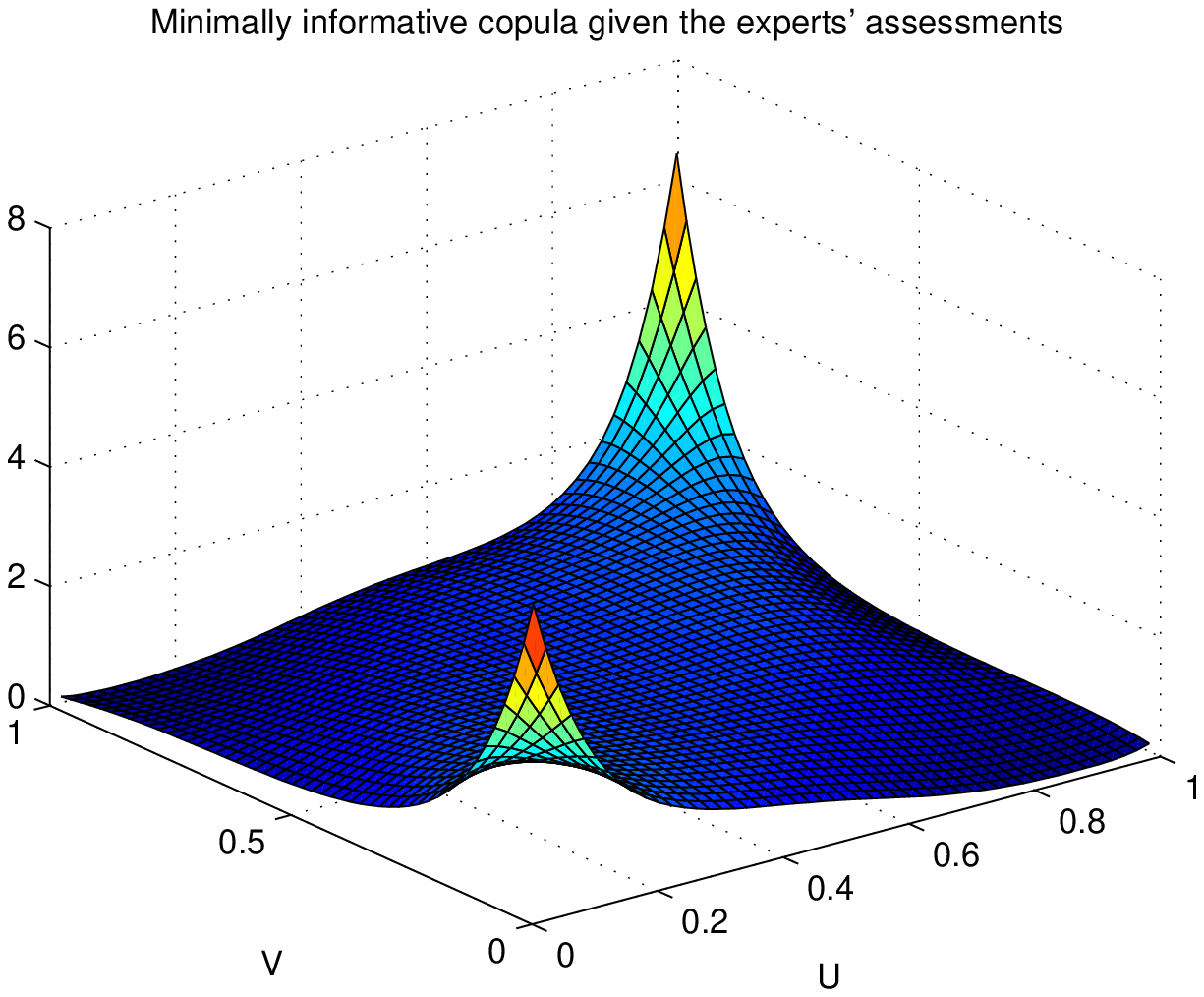}
\caption{The minimally informative copula between $M$ and $B$ using the orthonormal polynomial bases}
\label{M-Borthogonal}
\end{minipage}
\hspace{0.5cm}
\begin{minipage}[b]{0.45\linewidth}
\centering
\includegraphics[width=\textwidth]{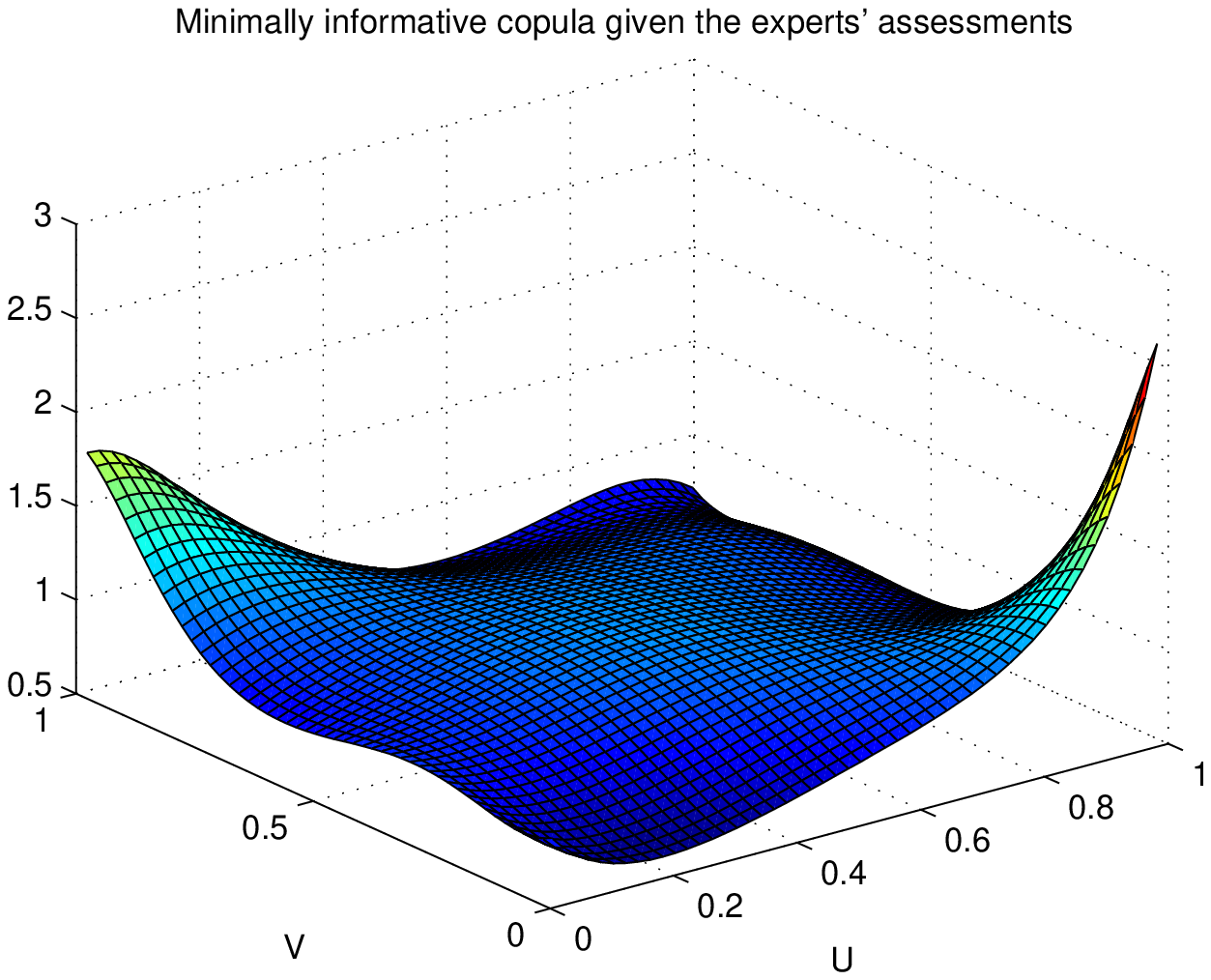}
\caption{The minimally informative copula between $M$ and $B$ using the orthonormal polynomial bases}
\label{B-Sorthogonal}
\end{minipage}
\end{figure}

The second copula in the first tree ($T_1$) is $C_{MB}$. Using the stepwise method, we choose the following orthonormal polynomial bases
\begin{eqnarray*}
h_1^{'} (M,B)=\varphi_1 (M) \varphi_1 (B),~~~ h_2^{'} (M,B)=\varphi_2 (M) \varphi_2 (B),~~~h_3^{'} (M,B)=\varphi_1 (M) \varphi_3 (B),\\
h_4^{'} (M,B)=\varphi_2 (M) \varphi_4 (B),~~~ h_5^{'} (M,B)=\varphi_4 (M) \varphi_1 (B),~~~h_6^{'} (M,B)=\varphi_1 (M) \varphi_5 (B)
\end{eqnarray*}
and we also select the following Legendre multiwavelets basis functions
\begin{eqnarray*}
h_1^{'} (M,B)=\phi^{1} (M) \phi^{1} (B),~~~ h_2^{'} (M,B)=\phi^{2} (M) \phi^{2} (B),~~~h_3^{'} (M,B)=\phi^{2} (M) \phi ^{4}(B),\\
h_4^{'} (M,B)=\phi^{3} (M) \phi^{1} (B),~~~ h_5^{'} (M,B)=\psi^{1} (M) \phi^{1} (B),~~~h_6^{'} (M,B)=\psi^{2} (M) \phi ^{1}(B)
\end{eqnarray*}
We similarly construct the minimally informative copulae associated with the orthonormal polynomial bases which is shown in Figure \ref{M-Borthogonal}.  Note that the minimum information copulas for the orthonormal polynomial and Legendre multiwavelets bases are quite similar, but the figure of later one to some extent is smoother than the former one. The constraints as the mean of the chosen orthonormal polynomial bases for the Norwegian Financial returns data are presented as
\begin{eqnarray*}
\alpha_{1}=0.4803 ,\alpha_{2}=0.2298 , \alpha_{3}=0.0841,
\alpha_{4}=0.0989 ,\alpha_{5}=0.0757 ,\alpha_{6}=-0.0112
\end{eqnarray*}
The parameter values associated with the fitted minimum information copula to the data with these constraints are given by
\begin{eqnarray*}
\lambda_1=0.5701,~~\lambda_2=0.0847,~~\lambda_3=0.0433,~~
\lambda_4=0.1000,~~\lambda_5=0.0830,~~\lambda_6=-0.0531
\end{eqnarray*}
The constraints for the Legendre multiwavelets bases are
\begin{eqnarray*}
\alpha_{1}=0.4803  ,\alpha_{2}=0.2298  , \alpha_{3}=0.0989 ,
\alpha_{4}=0.0757  ,\alpha_{5}=0.0531  ,\alpha_{6}=0.0463
\end{eqnarray*}
and the corresponding parameter values are as follows
\begin{eqnarray*}
\lambda_1=377.3642,~~\lambda_2=193.9254 ,~~\lambda_3=253.2358 ,~~
\lambda_4=281.7057,~~\lambda_5=-622.0234 ,~~\lambda_6=12.2802
\end{eqnarray*}
The log-likelihoods corresponding to the orthonormal polynomial and Legendre multiwavelets bases are 158.0013 and 159.72, respectively, which are again more than the log-likelihood calculated based on the ordinary polynomial bases.
\par
The third marginal copula is between $B$ and $S$. Similarly, the six bases are selected using the stepwise procedure, and the corresponding constraints and resulting Lagrange multipliers are given in Table  \ref{B-Sorthonormal} and Table \ref{B-Swavelet} for orthonormal and Legendre multiwavelets, respectively. The approximated minimally informative copula in terms of the orthonormal polynomial bases is shown in Figure \ref{B-Sorthogonal}. Note that the minimum information copula associated with the Legendre multiwavelets bases is very similar to the one given Figure \ref{B-Sorthogonal}, but to some extent is slightly smoother.
\begin{table}[ht]
\begin{minipage}[b]{0.45\linewidth}\centering
\begin{tabular}{|l|l|l|l|}
\hline
\multicolumn{1}{|c|}{Base}& \multicolumn{1}{|c|}{$\alpha_{i}$}& \multicolumn{1}{|c|}{$\lambda_{i}$} & \multicolumn{1}{|c|}{Log-}\\
\multicolumn{1}{|c|}{}& \multicolumn{1}{|c|}{}& \multicolumn{1}{|c|}{} & \multicolumn{1}{|c|}{Likelihood}\\
\hline
$\varphi_{1} (B) \varphi_{1} (S)$ &    -0.1557 &    -0.1467   &   \\
$\varphi_{2} (B) \varphi_{2} (S)$ &   0.1010 &    0.0836    &    \\
$\varphi_{3} (B) \varphi_{1} (S)$ &    -0.0510 &    -0.0426    & 20.13    \\
$\varphi_{1} (B) \varphi_{4} (S)$ &   -0.0378  &     -0.0365 &    \\
$\varphi_{2} (B) \varphi_{1} (S)$ &   0.0253  &    0.0257  &    \\
$\varphi_{5} (B) \varphi_{1} (S)$ &    0.0222 &    0.0240  &    \\ \hline
 \end{tabular}
\caption{The minimally informative copula for orthonormal polynomial bases between B and S }\label{B-Sorthonormal}
\end{minipage}
\hspace{0.5cm}
\begin{minipage}[b]{0.45\linewidth}
\centering
\begin{tabular}{|l|l|l|l|}
\hline
\multicolumn{1}{|c|}{Base}& \multicolumn{1}{|c|}{$\alpha_{i}$}& \multicolumn{1}{|c|}{$\lambda_{i}$} & \multicolumn{1}{|c|}{Log-}\\
\multicolumn{1}{|c|}{}& \multicolumn{1}{|c|}{}& \multicolumn{1}{|c|}{} & \multicolumn{1}{|c|}{Likelihood}\\
\hline
$\phi^{1} (B) \phi^{1} (S)$ &    0.4803 &    -70.35   &   \\
$\phi^{2} (B) \phi^{2} (S)$ &    0.2298 &    91.72    &    \\
$\psi^{2} (B) \phi^{2} (S)$ &    0.0539 &    16.61    &  25.07   \\
$\psi^{1} (B) \phi^{1} (S)$ &   0.0531  &     -22.29 &    \\
$\psi^{5} (B) \phi^{4} (S)$ &   0.0011  &    2.39  &    \\
$\psi^{4} (B) \phi^{2} (S)$ &    -0.0098 &    -3.49  &    \\ \hline
 \end{tabular}
\caption{The minimally informative copula for Legendre multiwavelets bases between B and S }\label{B-Swavelet}
\end{minipage}
\end{table}
\par
The conditional copulas in the second tree, $T_2$ can similarly be approximated using the minimum information approach. We only illustrate construction of the conditional minimum informative copula between $T|M$ and $B|M$, and the other conditional copulas in this tree can be similarly approximated. In order to calculate this copula, we divide the support of $M$ into some arbitrary sub-intervals or bins and then construct the conditional copula within each bin. To do so we select bases in the same way as for the marginal copulas and fit the copulae to the calculated mean values or constraints. Here, we use four bins so that the first copula is for $T,B|M\in (0,0.25)$. The bases for this copula based on the orthonormal polynomial basis are
\begin{eqnarray*}
h_1^{'} (T,B|M\in(0,0.25))=\varphi_2 (T) \varphi_1 (B),~~~  h_2^{'}(T,B|M\in(0,0.25))=\varphi_5 (T) \varphi_1 (B)\\
h_3^{'} (T,B|M\in (0,0.25))=\varphi_3 (T) \varphi_1 (B),~~~h_4^{'}(T,B|M\in(0,0.25))=\varphi_4 (T) \varphi_1 (B)\\
h_5^{'} (T,B|M\in(0,0.25))=\varphi_1 (T) \varphi_3 (B),~~h_6^{'} (T,B|M\in(0,0.25))=\varphi_2 (T) \varphi_3 (B)
\end{eqnarray*}
and the Legendre multiwavelets bases are also given by
\begin{eqnarray*}
h_1^{'} (T,B|M\in(0,0.25))=\phi^{1} (T) \phi^{1}(B),~~~  h_2^{'}(T,B|M\in(0,0.25))=\phi^{2} (T) \phi^{2} (B)\\
h_3^{'} (T,B|M\in (0,0.25))=\psi^{2} (T) \phi^{2} (B),~~~h_4^{'}(T,B|M\in(0,0.25))=\psi^{1} (T) \phi ^{1}(B)\\
h_5^{'} (T,B|M\in(0,0.25))=\psi^{5} (T) \phi^{4} (B),~~h_6^{'} (T,B|M\in(0,0.25))=\psi^{4} (T) \phi ^{2}(B)
\end{eqnarray*}
The mean values for orthonormal polynomial basis functions which will constrain the minimum information copula are
\begin{eqnarray*}
\alpha_{1}=-0.2995,~\alpha_{2}=-0.1240,~\alpha_{3}=-0.1634,~
\alpha_{4}=-0.0317,~\alpha_{5}=-0.0585,~\alpha_{6}=-0.0630
\end{eqnarray*}
and these expectations for Legendre multiwavelets bases are as follows
\begin{eqnarray*}
\alpha_{1}=0.4803,~\alpha_{2}=0.2298,~\alpha_{3}=0.0539,~
\alpha_{4}=0.0531,~\alpha_{5}=0.0011,~\alpha_{6}=-0.0098
\end{eqnarray*}
We can follow this process again for the remaining bins. Tables \ref{T-Borthonormal} and \ref{T-Bwavelet} show the mean values or constraints (denoted by $\alpha_{i}$) and corresponding Lagrange multipliers ($\lambda_{i}$) required to build the conditional minimum information copula between $T|M$ and $B|M$ for orthonormal polynomial and Legendre multiwavelets bases, respectively. The log-likelihood of the approximated copula in each bin is also reported in these tables.

\begin{table}
\begin{center}
\begin{tabular}{|l|l|l|l|l|}
\hline
 \multicolumn{1}{|c|}{Interval}&\multicolumn{1}{|c|}{Bases}& \multicolumn{1}{|c|}{$\alpha_{i}$} &\multicolumn{1}{|c|}{$\lambda_{i}$} &\multicolumn{1}{|c|}{Log-Likelihood}\\
\hline
      & $\varphi_2 (T)\varphi_1 (B)$ &  -0.2995 & -0.3511 &  \\
       & $\varphi_5 (T)\varphi_1 (B)$ &  -0.1240 & -0.135 &  \\
  $ 0<M<0.25$  &  $\varphi_3 (T)\varphi_1 (B)$ &  -0.1634 & -0.057 & 18.22 \\
     & $\varphi_4 (T)\varphi_1 (B)$ &  -0.0317 & 0.0776 & \\
     & $\varphi_1 (T)\varphi_3 (B)$ &    -0.0585 & -0.0705 & \\
      & $\varphi_2 (T)\varphi_3 (B)$ &      -0.0630 & 0.001 &  \\ \hline
   & $\varphi_3 (T)\varphi_1 (B)$ &  0.1504 & 0.1902 &  \\
      &  $\varphi_2 (T)\varphi_1 (B)$ &  0.0562 & 0.1051 &  \\
  $ 0.25<M<0.5$  &    $\varphi_4 (T)\varphi_1 (B)$ &  0.1030 & 0.1363 & 9.05 \\
      & $\varphi_1 (T)\varphi_4 (B)$ &  0.0836 & 0.0944 & \\
      & $\varphi_1 (T)\varphi_2 (B)$ &   0.0823 & 0.0804    & \\
   &$\varphi_4 (T)\varphi_2 (B)$ &      -0.0621 & -0.0094  &  \\ \hline
   &$\varphi_2 (T)\varphi_1 (B)$ &  0.1184 & 0.1679 &  \\
      & $\varphi_1 (T)\varphi_3 (B)$ &  -0.1080 & -0.2311 &  \\
 $ 0.5<M<0.75$  &    $\varphi_2 (T)\varphi_4 (B)$ &  0.0956 & 0.1459 & 9.74 \\
      & $\varphi_1 (T)\varphi_5 (B)$ &  -0.0815& -0.2047 & \\
     & $\varphi_1 (T)\varphi_2 (B)$ &   -0.0627 & -0.1869  & \\
      & $\varphi_3 (T)\varphi_1 (B)$ &      0.0245 & 0.1253  &  \\ \hline
   & $\varphi_1 (T)\varphi_1 (B)$ &  -0.2659 & -0.3177 &  \\
      & $\varphi_2 (T)\varphi_4 (B)$ &  0.1568 & 0.1135 &  \\
  $ 0.75<M<1$  &  $\varphi_4 (T)\varphi_1 (B)$ &  0.1025 & 0.1290 & 10.53\\
      & $\varphi_1 (T)\varphi_5 (B)$ &  -0.0079& 0.0526 & \\
      & $\varphi_1 (T)\varphi_3 (B)$ &   -0.1737 & -0.1007  & \\
      & $\varphi_3 (T)\varphi_3 (B)$ &  -0.0376 & 0.0456  &  \\ \hline
       \end{tabular}
\caption{Minimaly informative copula for orthonormal basis between $T$ and $B$ given
$M$}\label{T-Borthonormal}
\end{center}
\end{table}

\begin{table}
\begin{center}
\begin{tabular}{|l|l|l|l|l|}
\hline
 \multicolumn{1}{|c|}{Interval}&\multicolumn{1}{|c|}{Bases}& \multicolumn{1}{|c|}{$\alpha_{i}$} &\multicolumn{1}{|c|}{$\lambda_{i}$} &\multicolumn{1}{|c|}{Log-Likelihood}\\
\hline
      & $\phi^{2} (T)\phi^{1} (B)$ &  -0.020 & -2.53 &  \\
       & $\phi^{5} (T)\phi^{1} (B)$ &  -0.021 & 2.17 &  \\
  $ 0<M<0.25$  &  $\psi^{4} (T)\phi^{4} (B)$ &  -0.018 & -0.64 & 22.17 \\
     & $\phi^{4} (T)\phi^{5} (B)$ &  -0.002 & -3.00 & \\
     & $\psi^{0} (T)\phi^{6} (B)$ &    -0.004 & 2.35 & \\
      & $\psi^{5} (T)\phi^{4} (B)$ &      0.001 & 5.23 &  \\ \hline
   & $\phi^{3} (T)\phi^{1} (B)$ &  0.1504 & 0.226 &  \\
      &  $\psi^{4} (T)\phi^{4} (B)$ &  0.109 & -7.30 &  \\
  $ 0.25<M<0.5$  &    $\psi^{4} (T)\phi^{3} (B)$ &  0.102 & 9.97 & 12.08 \\
      & $\phi^{2} (T)\phi^{1} (B)$ &  0.056 & -3.31 & \\
      & $\phi^{4} (T)\phi^{1} (B)$ &   0.103 & -0.25    & \\
   &$\phi^{5} (T)\phi^{3} (B)$ &      0.106 & 10.46  &  \\ \hline
   &$\phi^{2} (T)\phi^{1} (B)$ &  0.118 & -269.93 &  \\
      & $\phi^{1} (T)\phi^{3} (B)$ &  -0.1080 & -439.13 &  \\
 $ 0.5<M<0.75$  &    $\psi^{1} (T)\phi^{2} (B)$ &  -0.102 & 104.95 & 11.30 \\
      & $\phi^{2} (T)\phi^{4} (B)$ &  0.096& -29.99 & \\
     & $\phi^{3} (T)\phi^{5} (B)$ &   0.093 & 373.59  & \\
      & $\psi^{4} (T)\phi^{5} (B)$ &      0.059 & -14.53  &  \\ \hline
   & $\phi^{1} (T)\phi^{1} (B)$ &  -0.2659 & 247.49 &  \\
      & $\phi^{2} (T)\phi^{4} (B)$ &  0.1568 & -110.67 &  \\
  $ 0.75<M<1$  &  $\phi^{4} (T)\phi^{1} (B)$ &  0.1025 & -108.01 & 12.86\\
      & $\psi^{1} (T)\phi^{4} (B)$ &   0.069& -222.75 & \\
      & $\phi^{3} (T)\phi^{5} (B)$ &    0.021 & -175.39  & \\
      & $\phi^{4} (T)\phi^{5} (B)$ &   0.063 & -15.69  &  \\ \hline
       \end{tabular}
\caption{Minimum information copula for Legendre multiwavelets between $T$ and $B$ given
$M$}\label{T-Bwavelet}
\end{center}
\end{table}

Note that the resulting minimum information copula over all bins for orthonormal polynomial bases is 47.54 and for Legendre multiwavelets is 58.41 while this amount for the ordinary polynomial bases is only 29.242 which indicates superiority of the former bases.
\par
We can obtain the conditional minimum informative copula in the third tree, $T_3$, similarly by dividing each of the conditioning variables' supports into four bins. Then the minimum information copulas for $T| (B,M)$ and $S|(B,M)$ are calculated on each combination of bins for $M$ and $B$ which makes 16 bins altogether for this tree. The bins, bases and log-likelihoods associated with each copula based on the orthonormal polynomial and Legendre multiwavelets basis are given in Tables \ref{T-Sorthonormal} and \ref{T-Swavelet}, respectively.
 \begin{table}[h]
\begin{center}
\begin{tabular}{|l|l|l|}
\hline
 \multicolumn{1}{|c|}{Interval}&\multicolumn{1}{|c|}{Bases}&\multicolumn{1}{|c|}{Log-Likelihood}\\
\hline
    $ 0<M<0.25$,  $ 0<B<0.25$  & $\varphi_1 \varphi_1, \varphi_3 \varphi_1,\varphi_1 \varphi_4,\varphi_5 \varphi_1,\varphi_4 \varphi_2,\varphi_1 \varphi_5$ &  8.93   \\ \hline
    $ 0<M<0.25$,   $ 0.25<B<0.5$ & $\varphi_1\varphi_1,\varphi_1 \varphi_3,\varphi_3 \varphi_3,\varphi_3 \varphi_2,\varphi_3 \varphi_1,\varphi_5 \varphi_1$ &  7.31   \\ \hline
    $ 0<M<0.25$,  $ 0.5<B<0.75$  &  $\varphi_2 \varphi_4,\varphi_3 \varphi_3,\varphi_5 \varphi_1,\varphi_2 \varphi_2,\varphi_1 \varphi_2,\varphi_3 \varphi_2$ &  6.81  \\ \hline
    $ 0<M<0.25$,   $ 0.75<B<1$ & $\varphi_4 \varphi_1,\varphi_2 \varphi_3,\varphi_1 \varphi_2,\varphi_1 \varphi_3,\varphi_2 \varphi_2,\varphi_2 \varphi_4$ &  9.65 \\ \hline
    $ 0.25<M<0.5$,   $ 0<B<0.25$ & $\varphi_1 \varphi_1,\varphi_2 \varphi_4,\varphi_3 \varphi_1,\varphi_3 \varphi_3,\varphi_1 \varphi_5,\varphi_1 \varphi_2 $ &    8.63  \\ \hline
    $ 0.25<M<0.5$,   $ 0.25<B<0.5$ &$\varphi_1\varphi_1,\varphi_5 \varphi_1,\varphi_2 \varphi_1,\varphi_4 \varphi_1,\varphi_3 \varphi_1,\varphi_1 \varphi_4$ &     7.67   \\ \hline
    $ 0.25<M<0.5$,   $ 0.5<B<0.75$ &$\varphi_1\varphi_1,\varphi_2 \varphi_1,\varphi_5 \varphi_1,\varphi_2 \varphi_3,\varphi_1 \varphi_5,\varphi_3 \varphi_2 $ &  9.5   \\ \hline
    $ 0.25<M<0.5$,   $ 0.75<B<1$ & $\varphi_3 \varphi_2,\varphi_1 \varphi_3,\varphi_2 \varphi_4,\varphi_3 \varphi_3,\varphi_3 \varphi_1,\varphi_4 \varphi_2 $ &  5.62   \\ \hline
    $ 0.5<M<0.75$,$ 0<B<0.25$ & $\varphi_1\varphi_1,\varphi_3 \varphi_2,\varphi_2 \varphi_4,\varphi_1 \varphi_3,\varphi_1 \varphi_5,\varphi_4 \varphi_2$ &  4.93  \\ \hline
    $ 0.5<M<0.75$,   $ 0.25<B<0.5$ & $\varphi_1 \varphi_1,\varphi_1 \varphi_2,\varphi_2 \varphi_1,\varphi_1 \varphi_3,\varphi_3 \varphi_1,\varphi_2 \varphi_2$ &  10.49   \\ \hline
    $ 0.5<M<0.75$,   $ 0.5<B<0.75$& $\varphi_1 \varphi_1 ,\varphi_1 \varphi_2,\varphi_4 \varphi_1,\varphi_3 \varphi_3,\varphi_2 \varphi_2,\varphi_5 \varphi_1$ &   8.97  \\ \hline
    $ 0.5<M<0.75 $,  $ 0.75<B<1$ &$\varphi_1 \varphi_1,\varphi_3 \varphi_3,\varphi_4 \varphi_1,\varphi_2 \varphi_3,\varphi_1 \varphi_4,\varphi_2 \varphi_4$ &      10.08   \\ \hline
    $ 0.75<M<1 $,   $ 0<B<0.25$ &$\varphi_4 \varphi_2,\varphi_5 \varphi_1,\varphi_1 \varphi_5,\varphi_3 \varphi_2,\varphi_1 \varphi_2,\varphi_1 \varphi_4 $ &  3.7  \\ \hline
    $ 0.75<M<1 $,   $ 0.25<B<0.5$ &$\varphi_2 \varphi_2,\varphi_2 \varphi_4,\varphi_2 \varphi_3,\varphi_1 \varphi_1,\varphi_4 \varphi_1,\varphi_4 \varphi_2 $ &  8.7   \\ \hline
    $ 0.75<M<1 $, $ 0.5<B<0.75$ &    $\varphi_1 \varphi_4,\varphi_1 \varphi_1,\varphi_5 \varphi_1,\varphi_1 \varphi_3,\varphi_3 \varphi_2,\varphi_3 \varphi_3 $ &  5.61  \\ \hline
    $ 0.75<M<1 $,    $ 0.75<B<1$& $\varphi_2 \varphi_2 ,\varphi_1 \varphi_1,\varphi_2 \varphi_1,\varphi_1 \varphi_5,\varphi_4 \varphi_1,\varphi_3 \varphi_2$ &      20.24  \\ \hline

 \end{tabular}
\caption{Minimum information copula for orthonormal basis between $T$ and $S$ given$M$  and $B$}\label{T-Sorthonormal}
\end{center}
\end{table}

 \begin{table}[h]
\begin{center}
\begin{tabular}{|l|l|l|}
\hline
 \multicolumn{1}{|c|}{Interval}&\multicolumn{1}{|c|}{Bases}&\multicolumn{1}{|c|}{Log-Likelihood}\\
\hline
     $ 0<M<0.25$,  $ 0<B<0.25$  & $\phi^{1} \phi^{2}, \phi^{3} \phi^{1},\psi^{0} \phi^{2},\phi^{4} \phi^{5},\phi^{5} \phi^{4},\psi^{4} \phi^{2}$ &  10.64   \\ \hline
    $ 0<M<0.25$,   $ 0.25<B<0.5$ & $\phi^{1}\phi^{1},\psi^{2} \psi^{1},\phi^{1} \phi^{3},\phi^{3} \phi^{3},\psi^{1} \phi^{5},\psi^{5} \phi^{5}$ &  9.42   \\ \hline
    $ 0<M<0.25$,  $ 0.5<B<0.75$  &  $\psi^{0} \phi^{5},\phi^{5} \phi^{3},\phi^{3} \phi^{3},\phi^{1} \phi^{3},\phi^{5} \phi^{1},\phi^{2} \phi^{4}$ &  13.67  \\ \hline
    $ 0<M<0.25$,   $ 0.75<B<1$ & $\phi^{4} \phi^{1},\phi^{2} \phi^{3},\phi^{1} \phi^{2},\phi^{1} \phi^{3},\psi^{3} \phi^{5},\phi^{5} \phi^{2}$ &  10.42 \\ \hline
    $ 0.25<M<0.5$,   $ 0<B<0.25$ & $\phi^{1} \phi^{1},\phi^{2} \phi^{4},\phi^{2} \phi^{5},\phi^{5} \phi^{4},\psi^{3} \phi^{2},\phi^{5} \phi^{3} $ &  12.99  \\ \hline
    $ 0.25<M<0.5$,   $ 0.25<B<0.5$ &$\psi^{3}\phi^{1},\psi^{2} \phi^{5},\psi^{2} \phi^{3},\phi^{2} \phi^{5},\phi^{1} \phi^{2},\psi^{3} \phi^{2}$ &  15.67 \\ \hline
    $ 0.25<M<0.5$,   $ 0.5<B<0.75$ &$\phi^{1}\phi^{1},\phi^{5} \phi^{2},\phi^{2} \phi^{1},\phi^{4} \phi^{4},\psi^{2} \phi^{1},\psi^{3} \phi^{4} $ &  10.56   \\ \hline
    $ 0.25<M<0.5$,   $ 0.75<B<1$ & $\phi^{5} \phi^{2},\psi^{0} \phi^{3},\phi^{1} \phi^{1},\phi^{5} \phi^{4},\psi^{0} \phi^{5},\phi^{3} \phi^{1} $ &  10.77 \\ \hline
    $ 0.5<M<0.75$,$ 0<B<0.25$ & $\phi^{3}\phi^{2},\psi^{5} \phi^{5},\phi^{1} \phi^{4},\psi^{4} \phi^{2},\psi^{4} \phi^{1},\phi^{3} \phi^{3}$ &  9.89  \\ \hline
    $ 0.5<M<0.75$,   $ 0.25<B<0.5$ & $\phi^{5} \phi^{2},\phi^{1} \phi^{1},\psi^{3} \phi^{1},\psi^{2} \phi^{1},\phi^{2} \phi^{2},\phi^{4} \phi^{4}$ &10.26\\ \hline
    $ 0.5<M<0.75$,   $ 0.5<B<0.75$& $\phi^{1} \phi^{1} ,\phi^{2} \phi^{5},\psi^{3} \phi^{3},\psi^{3} \phi^{5},\psi^{0} \phi^{4},\phi^{3} \phi^{3}$ & 14.01\\ \hline
    $ 0.5<M<0.75 $,  $ 0.75<B<1$ &$\phi^{3} \phi^{4},\psi^{0} \phi^{4},\phi^{1} \phi^{1},\phi^{3} \phi^{5},\phi^{3} \phi^{3},\psi^{3} \phi^{2}$ &17.97 \\ \hline
    $ 0.75<M<1 $,   $ 0<B<0.25$ &$\phi^{1} \phi^{1},\phi^{2} \phi^{3},\psi^{4} \phi^{1},\phi^{3} \phi^{3},\psi^{3} \phi^{3},\psi^{2} \phi^{1} $ &  11.17  \\ \hline
    $ 0.75<M<1 $,   $ 0.25<B<0.5$ &$\phi^{1} \phi^{3},\phi^{4} \phi^{2},\phi^{5} \phi^{1},\psi^{4} \phi^{1},\psi^{2} \phi^{2},\psi^{3} \phi^{1} $ &  14.31   \\ \hline
    $ 0.75<M<1 $, $ 0.5<B<0.75$ &    $\phi^{1} \phi^{1},\phi^{2} \phi^{5},\psi^{0} \phi^{2},\psi^{1} \phi^{4},\psi^{3} \phi^{3},\phi^{4} \phi^{4} $ &  10.61\\ \hline
    $ 0.75<M<1 $,    $ 0.75<B<1$& $\phi^{3} \phi^{1} ,\phi^{2} \phi^{2},\phi^{1} \phi^{5},\psi^{4} \phi^{1},\phi^{2} \phi^{4},\phi^{5} \phi^{5}$ &  24.39  \\ \hline
 \end{tabular}
\caption{Minimally informative copula for Legendre multiwavelets between $T$ and $S$ given$M$ and $B$}\label{T-Swavelet}
\end{center}
\end{table}
Thus the log-likelihood of the overall vine, obtained by summing the log-likelihoods of each of
the component copulas above, is 388.859.

The log-likelihood of the overall pair-copula model using the orthonormal polynomial (and Legendre multiwavelets) bases, derived by adding the log-likelihoods of the copulas constructed above, is then 434.135 (and this amount for Legendre multiwavelets is 552.25). These values are considerably greater than the log-likelihoods of the fitted pair-copula models to the data using the Gaussian copula, $t$-copula and the approximated pair-copula model using the ordinary polynomial bases.

\subsection{Comparison To Other Approaches}
In this subsection, we compare our method with the other methods used to approximate the multivariate distribution fitted to the Norwegian financial returns data. In order to make a comparison we compute the log-likelihood of the approximated density function by the method presented in this paper and other approaches reported in Aas et. (2009) and Bedford et. (2012). The log-likelihood of the overall pair-copula model using the orthonormal polynomial and Legendre multiwavelets bases, obtained by adding the log-likelihoods of each of the component copulas presented above, are 434.135 and 552.25, respectively. These values are much greater than that obtained using the $t$-copula examined by Aas et al (2009) of 291.801 and the minimum information copula based on the ordinary polynomial bases of Bedford et al (2012) of 388.859. Note that, if we only use five bases to approximate the multivariate density of interest, the log-likelihoods associated with orthonormal polynomial and Legendre multiwavelets bases will be 429.3982 and 446.235, respectively, which are still clearly better than the model proposed by Bedford et al (2012) based on the six ordinary polynomial bases. We have computed the log-likelihood of the data sample for five different copula models used on the same vine structure: The Gaussian copula, the $t$-copula used by Aas et al. (2009), the minimum information copula using the ordinary polynomial bases presented by Bedford et al. (2012) and our approximated copulas. We illustrate the corresponding results in Table \ref{Comparison}.
\begin{table}
\begin{center}
\begin{tabular}{|l|l|}
\hline
 \multicolumn{1}{|c|}{Type of copula}&\multicolumn{1}{|c|}{Log-Likelihood}\\
\hline
   Gaussian copula (Aas et al. 2009) & 263.5052    \\ \hline
   $t$ copula (Aas et al. 2009) &    291.8014 \\ \hline
   Minimum information copula based  &   388.859   \\
    on polynomial basis (Bedford et al., 2012)  &          \\ \hline
    Minimum information copula &  434.135   \\
    based on orthonormal polynomial &   \\    \hline
    Minimum information copula  &  552.25   \\
    based on Legendre multiwavelets &   \\  \hline
    \end{tabular}
\caption{Comparison between different models.}\label{Comparison}
\end{center}
\end{table}

\section{Conclusion}
In this paper, we extend the novel method originally presented by Bedford et al (2012) to approximate a multivariate distribution by any vine structure to any degree of approximation. The main idea to implement this approximation method is to use  the minimum information copulae that can be determined to any required degree of precision based on the   data available. To approximate a multivariate distribution by this method, we need to specify: 1) a vine structure; 2) a basis family; 3) for each part of vine, expected values for the certain functions associated with some constraints on each pairwise copula.
\par
Bedford et al (2012) approximate all conditional copulas using linear combinations of the ordinary polynomial basis functions. We make this approximation more precise by choosing more appropriate basis family. We concentrate on the orthonormal polynomial basis functions and Legendre multiwavelets in this paper. The Legendre multiwavelets and orthonormal polynomial basis functions are shown that to be more convenient than some other natural basis for the purpose of calculation. A very nice property of the orthonormal polynomial basis is that adding a new item to the expansion does not change
coefficient of the already found shorter expansion which is not the case for the non-orthonormal basis where any new item has in general nonzero projection on previous items. It means that the already found coefficients of the expansion would have to be changed. The Legendre multiwavelets basis, not only has this property, but the computation of the minimum information copula using this basis becomes even faster and the approximation would considerably improve. In other words, applying these basis is so important from three main aspects: firstly, less computation time is required to approximate the minimum information copula of interest; secondly, the fitted models to the data using the minimum information copulas based on the orthonormal polynomial and Legendre multiwavelets  bases are better in the sense that their log-likelihoods are much larger than than log-likelihood of the alternative models; thirdly, the approximations made in this paper are robust in the sense that they are not sensitive to the initial values chosen for the parameter values.
\par
In addition to these properties, our method has this property that it can be used to build arbitrarily good approximations to the original distribution. One of the most clear sources of potential error in our approximation is the choice of base where it is convenient to take a low number of functions $h_{i}$. The terms chosen in both orthonormal polynomial and Legendre multiwavelets would generate asymmetric copulas which seems to have great impact in modelling general data sets. The use of large numbers of functions does give more accuracy, at the cost of considerable extra computation at the construction stage but at no extra cost at the sampling stage. Indeed, we can approximate the requested model more precisely using less numbers of basis functions proposed in this paper and with smaller computation time than the alternative methods. In fact, the generalization made in this paper gives natural ways to generate asymmetric copulas, and simple ways to specify non-constant conditional correlations (or other moments). At moment, we are investigating some alternative methods to the stepwise method used in this paper to find the most optimal basis functions in a sense that
with smaller number of these bases, we would get the largest
log-likelihood.
\par
The method used in this paper is very flexible and any functions can be used to construct the minimum information copulas used here. This method can be use for modeling more complex applications at which basis functions should be computed in computer codes. Due to numerous evaluation of these function to construct the minimum information distribution, the computation and then approximation will be infeasible. One suggestion to ease the computation and reduce the complexity of model is to use the Gaussian process emulators.\\
\par
\textbf{Acknowledgement}: The authors are grateful to Professor Tim Bedford for his helpful comments for some parts of the paper.


\begin{thebibliography}{28}
\bibitem{Aas-etal09} Aas, K., Czado, K. C., Frigessi, A., and Bakken, H. (2009). Pair-copula constructions of multiple dependence.
\textit{Insurance, Mathematics and Economics}, \textbf{44}, 182--198.
\bibitem{pa} Alpert, B., Beylkin, G., Gines, D., Vozovoi, L. (2002). Adaptive solution of partial differential equations in multiwavelet bases. \emph{J. Comput. Phys.}, \textbf{182}: 149-190.
\bibitem{pa} Bedford, T. (2006). Interactive expert assignment of minimally-informative copulae, \emph{Management Science Working Paper} No. 5.
\bibitem{Bedford-Cooke01} Bedford, T. and Cooke. R. M. (2001). Probability density decomposition for conditionally dependent random variables modeled by vines. \emph{Annals of Mathematics and Artificial Intelligence} \textbf{32}, 245--268.
\bibitem{pa} Bedford, T., and Cooke. R. M. (2002). Vines - a new graphical model for dependent random
variables. \emph{Annals of Statistics}, \textbf{30(4)}: 1031--1068.
random variables modeled by vines. \emph{Annals of Mathematics and Artificial Intelligence} \textbf{32}, 245--268.
\bibitem{Bedford-Danesh10} Bedford, T. and Daneshkhah, A. (2010). Approximating Multivariate Distributions with Vines. \emph{Submitted to Operation Research}.
\bibitem{Bedford et al 12} Bedford, T. and Daneshkhah, A., Wilson, K. (2012). Approximate Uncertainty Modeling with Vine copulas. \emph{Submitted to European Journal of Operation Research}.
\bibitem{Bedford-Meeuwissen97} Bedford, T., and Meeuwissen, A. (1997). Minimally informative distributions with given
rank correlation for use in uncertainty analysis. \emph{Journal of Statistical Computation and Simulation}, \textbf{57(1 - 4)}: 143 - 174.
\bibitem{pa} Bollerslev, T.  (1986). Generalized Autoregressive Conditional Heteroskedasticity. \textit{ Journal of Econometrics }, \textbf{31}, 307--327.
\bibitem{Borwein-etal94} Borwein, J., Lewis, A., and Nussbaum, R. (1994). Entropy minimization, DAD problems, and doubly stochastic kernels. \emph{Journal of Functional Analysis}, \textbf{123}, 264-307.
\bibitem{pa} Embrechts, P., F. Lindskog, and A. J. McNeil. (2003). Modelling Dependence with Copulas and Applications to Risk Management. In Handbook of Heavy Tailed Distributions in Finance. \textit{ Amsterdam: Elsevier/North-Holland}, \textbf{31}, 307--327.
\bibitem{pa} Engle, R. F.  (1982). Autoregressive conditional heteroscedasticity with estimates of the variance of United
Kingdom. \textit{ Econometrica}, \textbf{50(4)}, 987--1007.
\bibitem{pa} Gui, W. (2009). \emph{Adaptive Series Estimators for Copula Densities}. PhD thesis, Florida State University College of Arte and Sciences.
\bibitem{Koe97} Joe, H. (1997). \emph{Multivariate Models and Dependence Concepts}. Chapman \& Hall, London.
\bibitem{pa} Kurowicka, D., and Cooke. R. (2006). \emph{Uncertainty Analysis with High Dimensional Dependence Modelling}. John Wiley.
\bibitem{Koe97} Kurowicka, D. and Joe, H. (2011). \emph{Dependence Modeling: Vine Copula Handbook}. World Scientific, Singapore.
\bibitem{pa} Lagarias, J. C., Reeds, J. A., Wright, M. H., and  Wright. P. E. (1998). Convergence properties of the Nelder-Mead simplex method in low dimensions. \emph{SIAM Journal of Optimization}, \textbf{9(1)}: 112--147.
\bibitem{pa} Lewandowski, D. (2008).\emph{High Dimensional Dependence: Copulae, Sensitivity, Sampling}. PhD thesis, Delft University.
\bibitem{pa} Ljung, G. M., and Box, G. E. P. (1978). On a Measure of lack of Fit in Time Series Models. \textit{Biometrika}, \textbf{65(2)}, 297--303.
\bibitem{pa} Locke, B., and Peter, A. (2012). Multiwavelet Density Estimation, submitted to \emph{Computational Statistics and Data Analysis}.
\bibitem{pa} Shamsi, M.,  Razzaghi, M. (2005). Solution of Hallen's Integral Equation Using Multiwavelets,. \emph{Computer Physics Communications}, \textbf{168}: 187-197.
\end{thebibliography}
\end{document}